\documentclass[camera,letterpaper,nomarginnotes,nonarrowgutter]{jpaper}

\usepackage{amsmath, amssymb, amsfonts}
\usepackage{graphicx}
\usepackage{textcomp}
\usepackage{url}
\usepackage{algorithm}
\usepackage{algpseudocode}
\usepackage{multirow}
\usepackage{tikz}
\usepackage{marginnote}
\usepackage{fancyhdr}
\usepackage{titlesec}
\usepackage{dblfloatfix}
\usepackage{makecell}
\usepackage{comment}
\usepackage{xspace}
\usepackage{xcolor, soul}
\usepackage{enumitem}
\usepackage{tabularx} % For extended table functionalities
\PassOptionsToPackage{bookmarks=true,breaklinks=true,letterpaper=true,colorlinks,citecolor=blue,linkcolor=blue,urlcolor=blue}{hyperref}
\usepackage{hyperref}
\usepackage{setspace}
\usepackage[colorinlistoftodos,prependcaption,textsize=scriptsize]{todonotes} % For margin notes
\usepackage[sort,compress]{cite}

\usepackage{siunitx}
\usepackage{datetime}

\definecolor{bleudefrance}{rgb}{0.19, 0.55, 0.91}
\definecolor{caribbeangreen}{rgb}{0.0, 0.8, 0.6}
\definecolor{english}{rgb}{0.0, 0.5, 0.0}
\definecolor{teal}{rgb}{0.0, 0.5, 0.5}
\definecolor{tyrianpurple}{rgb}{0.4, 0.01, 0.24}
\definecolor{violet}{rgb}{0.56, 0.0, 1.0}
\definecolor{darkorange}{rgb}{0.78, 0.43, 0}

\newcommand{\fig}[1]{{Fig.#1}} %

\newcommand{\head}[1]{{\noindent\textbf{#1.}}} %

\newcommand*\circled[1]{\tikz[baseline=(char.base)]{
            \node[shape=circle,draw,inner sep=0.4pt,fill=black, text=white] (char) {#1};}}

\newcommand*\circleds[1]{\tikz[baseline=(char.base)]{
            \node[shape=circle,draw,inner sep=-0.2pt,fill=black, text=white] (char) {#1};}}

\newcommand{\isp}{{ISP}}

\newcommand{\ideal}{{IFP (Ideal)}}
\newcommand{\cpu}{{CPU}}

\newcommand{\venice}{{IFP (Venice)}}
\newcommand{\baseline}{{IFP (Shared-Channel SSD)}}
\newcommand{\namePaper}{{Lattice}} %TITLE

\newcommand{\squishlist}{
 \begin{list}{$\circ$}
  { \setlength{\itemsep}{0pt}
     \setlength{\parsep}{0pt}
     \setlength{\topsep}{3pt}
     \setlength{\partopsep}{0pt}
     \setlength{\leftmargin}{1em}
     \setlength{\labelwidth}{1em}
     \setlength{\labelsep}{0.5em} } }

\newcommand{\squishend}{  \end{list}  }

\def\BibTeX{{\rm B\kern-.05em{\sc i\kern-.025em b}\kern-.08em
    T\kern-.1667em\lower.7ex\hbox{E}\kern-.125emX}}

\fancypagestyle{firstpage}{%
  \fancyhf{}
  \fancyfoot[C]{\thepage}
  
}

\title{\namePaper{}: Alleviating the Shared-Channel Bottleneck \\ in Solid-State Drives to Enable Scalable In-Flash Processing}
\author{
    Rakesh Nadig\textsuperscript{$\dagger$} \hspace{0.5em} Manos Frouzakis\textsuperscript{$\dagger$} \hspace{0.5em} 
    Nika Mansouri Ghiasi\textsuperscript{$\dagger$} \hspace{0.5em} 
    Yonggon Park\textsuperscript{$\dagger$$\ddagger$} \vspace{0.2em} \\
    Mayank Kabra\textsuperscript{$\dagger$} \hspace{0.5em} 
    Sahand Divsalar\textsuperscript{$\dagger$} \hspace{0.5em} 
    Harshita Gupta\textsuperscript{$\dagger$} \hspace{0.5em} 
    Andreas Kosmas Kakolyris\textsuperscript{$\dagger$} \vspace{0.2em} \\ \hspace{0.5em} 
    Jisung Park\textsuperscript{$\ddagger$} \hspace{0.5em} 
    Mohammad Sadrosadati\textsuperscript{$\dagger$} \hspace{0.5em} 
    Onur Mutlu\textsuperscript{$\dagger$} \vspace{0.5em} \\
    \textsuperscript{$\dagger$}\emph{ETH Zürich} \hspace{0.5em} \textsuperscript{$\ddagger$}\emph{POSTECH}
}

\begin{document}
\maketitle
\thispagestyle{firstpage}

%%%%%% -- PAPER CONTENT STARTS-- %%%%%%%%
\begin{abstract}
Data-intensive applications spend significant time and energy moving large volumes of data between storage, memory, and processing units.
In-flash processing (IFP) reduces these costs by performing computations inside NAND flash chips, where large datasets already reside.
To exploit the flash array's massive parallelism, prior IFP techniques distribute computation across many flash chips and aggregate the partial results they produce.
However, two architectural bottlenecks limit IFP's scalability in a solid-state drive (SSD). 
First, an SSD's shared-channel organization serializes data transfers and exposes only a fraction of the SSD's internal bandwidth. 
While connecting more flash chips to a channel increases compute parallelism, the channel bandwidth remains fixed.
Second, the SSD controller orchestrates all data transfers and computation. 
Every flash chip's partial result passes through the SSD controller before aggregation, increasing aggregation latency and energy.
Prior network-in-SSD architectures propose an interconnection network of flash chips to provide more paths and reduce serialized data transfers.
Unfortunately, 
they still rely on controller-orchestrated data transfers and lack in-network aggregation. 

Our goal is to scale distributed IFP by enabling flash chips to exchange and aggregate partial results directly.
To this end, we propose \textit{\textbf{\namePaper{}}}, a lightweight IFP architecture based on three key ideas.
First, to enable direct chip-to-chip communication, \namePaper{} embeds an IFP-aware router adjacent to each flash chip and connects neighboring routers via short, high-speed links.
Second, to coordinate distributed computation efficiently, \namePaper{} introduces IFP-aware communication primitives that support operand distribution, synchronization, and in-network aggregation of partial results.
Third, to reduce data movement, \namePaper{} performs complex computation (e.g., matrix multiplication) across flash chips and simple non-linear operations (e.g., masking) in the routers during data transfer.
We demonstrate \namePaper{}'s benefits on machine learning (ML) workloads by reformulating general matrix-matrix multiplication (GEMM) and convolution operations as distributed vector-scalar multiplications, which map efficiently to IFP primitives.

Across five GEMM- (convolution-) intensive ML workloads, \namePaper{} outperforms IFP on a state-of-the-art network-in-SSD architecture by 2.6$\times$ (2.5$\times$) and reduces energy consumption by 1.3$\times$ (1.7$\times$) on average.
\namePaper{}'s benefits come at low area and power overheads: each router chip adds printed circuit board area equal to 8\% of a flash die's footprint and consumes 2.2 mW, and each link consumes 1.08 mW during a 4 KB page transfer.
\end{abstract}
\section{Introduction} 
Data-intensive workloads (e.g., databases~\cite{doty1980magnetic, elmasri2007fundamentals, kocberber2013meet, augusta-sigmod-2015, Oracle, idreos2012monetdb, wu2014q100, chan-sigmod-1998, oneil-ideas-2007, li-vldb-2014, li-sigmod-2013, goodwin-SIGIR-2017, seshadri-micro-2013, seshadri2017ambit, seshadri-ieeecal-2015, hajinazar2021simdram}, web search~\cite{ccetin2016blind, pugsley2014ndc, cho-wondp-2013}, data analytics~\cite{jia2017understanding, jun-isca-2015}, genomics~\cite{alser-bioinformatics-2017, loving-bioinformatics-2014, xin-bioinformatics-2015, cali-micro-2020, kim-genomics-2018, lander-nature-2001, altschul-jmb-1990, myers-jacm-1999, ghiasi2022genstore, ghiasi_megis_2024, ghiasi2026sage, ghiasi2026grains}, graph processing~\cite{shim2022gp3d, zhang2018graphp, ahn-isca-2015, zhuo-micro-2019, huang2020heterogeneous, song2018graphr, dai-tcad-2019}, machine learning (ML)~\cite{ankit2019puma, low.vldb12, wang2020survey, sheng2023flexgen,pan2024instattention,jang2025inf, lincoln-hpca, zhang2023g10, yuan2025teraio}, cryptography~\cite{tuyls-springer-2005, manavski-spcom-2007, nejatollahi2020cryptopim, han-spie-1999}) spend a significant fraction of their execution time and energy moving large volumes of data between storage, memory, and host computation units~\cite{googleWorkloads, keckler2011gpus, ghiasi2022alp, abulila2019flatflash}.
Storage-centric computing techniques (e.g.,~\cite{
park2016storage, kabra2025ciphermatch, mailthody-micro-2019, kim-fast-2021, kang-msst-2013, torabzadehkashi-pdp-2019, wang-eurosys-2019,acharya-asplos-1998,keeton-sigmod-1998, tiwari-fast-2013, tiwari-hotpower-2012, boboila-msst-2012, bae-cikm-2013,torabzadehkashi-ipdpsw-2018, pei-tos-2019, do-sigmod-2013, kim-infosci-2016, riedel-computer-2001,riedel-vldb-1998, liang-atc-2019,cho-wondp-2013, jun-isca-2015, lee-ieeecal-2020, ajdari-hpca-2019,liang-fpl-2019,jun-hpec-2016, kim-sigops-2020, lee2022smartsage, li2023ecssd, ruan2019insider, wang2016ssd1, li-atc-2021, jeong-tpds-2019, gouk2024dockerssd, ghiasi_megis_2024, ghiasi2022genstore, kang-micro-2021, yavits2021giraf, kim2023optimstore, lim-icce-2021, narasimhamurthy2019sage, jun-isca-2018, fakhry2023review, gu-isca-2016, yang2023lambda, jo2016yoursql, chen2025reis, seshadri-osdi-2014, mahapatra2025rag, pan2024instattention, lee2017extrav, wang2024beacongnn, yu2024cambricon, pan2024instinfer, lincoln-hpca, jang2025inf, jaliminchecs, kang2024isp, wang-damon-2016, koo-micro-2017, kang-tc-2021, park2022flash, gao-micro-2021, chen2024aresflash, soysal2025mars, nadig2026conduit, kim2025crossbit}) move computation closer to data in storage devices, e.g., solid-state drives (SSDs).
In-storage processing (ISP) (e.g.,~\cite{park2016storage,   kabra2025ciphermatch, mailthody-micro-2019, kim-fast-2021, kang-msst-2013, torabzadehkashi-pdp-2019, seshadri-osdi-2014, wang-eurosys-2019,acharya-asplos-1998,keeton-sigmod-1998, wang-damon-2016, koo-micro-2017, tiwari-fast-2013, tiwari-hotpower-2012, boboila-msst-2012, bae-cikm-2013,torabzadehkashi-ipdpsw-2018, pei-tos-2019, do-sigmod-2013, kim-infosci-2016, riedel-computer-2001,riedel-vldb-1998, liang-atc-2019,cho-wondp-2013, jun-isca-2015, lee-ieeecal-2020, ajdari-hpca-2019,liang-fpl-2019,jun-hpec-2016, kim-sigops-2020, lee2022smartsage, li2023ecssd, ruan2019insider, wang2016ssd1, li-atc-2021, jeong-tpds-2019, gouk2024dockerssd, ghiasi_megis_2024, ghiasi2022genstore, kang-micro-2021, yavits2021giraf, kim2023optimstore, lim-icce-2021, narasimhamurthy2019sage, jun-isca-2018, fakhry2023review, gu-isca-2016, yang2023lambda, jo2016yoursql, chen2025reis, nadig2026conduit}) performs computations on embedded cores or an FPGA near the SSD controller (e.g., SmartSSD~\cite{lee-ieeecal-2020}), but operands still move through flash channels.  
In-flash processing (IFP) (e.g.,~\cite{park2022flash, gao-micro-2021, chen2024aresflash, kabra2025ciphermatch, kim2025crossbit,kang-tc-2021}) reduces this traffic by exploiting the operating principles of memory cells and peripheral circuitry of a NAND flash chip to perform bitwise and arithmetic operations.

Prior IFP techniques (e.g.,~\cite{park2022flash, gao-micro-2021, kabra2025ciphermatch, chen2024aresflash, kim2025crossbit, kang-tc-2021, chen2024search, nadig2026conduit}) partition data across flash chips to perform computations in parallel.
This data partitioning requires frequent synchronization and aggregation of intermediate results, which exposes two architectural bottlenecks that limit IFP's scalability, performance, and energy efficiency.
First, although flash chips can perform computations concurrently, a modern SSD's shared-channel organization exposes \emph{only} a fraction of the SSD's internal bandwidth.
In our illustrative SSD configuration with 64 flash chips, the flash chips can concurrently read at up to 80 GiB/s (for SLC reads), but shared channels deliver only 9.6 GiB/s of total bandwidth. 
This bandwidth mismatch forces flash chips to wait for channel availability~\cite{nadig2023venice, tavakkol2012network}, serializing data transfers. 
This communication bottleneck worsens as more flash chips are connected to a shared channel.
Second, every inter-chip exchange of intermediate results incurs controller round-trips. 
The SSD controller orchestrates all computations, data transfers, and aggregation operations during IFP (see \S\ref{subsec:motivation_shared_channel}).
The extra transfers increase aggregation latency and energy consumption. 
As more chips exchange partial results, controller traffic grows.

To overcome these architectural bottlenecks, prior network-in-SSD techniques (e.g.,~\cite{nadig2023venice, kim2022networked, tavakkol2012network, cui2024smartnetssd}) employ an interconnection network of flash chips inside the SSD to provide path diversity and reduce serialized I/O data transfers.
Our motivational analysis (see \S\ref{sec:motivation}) shows that IFP on a state-of-the-art network-in-SSD architecture~\cite{nadig2023venice} falls short of IFP on an ideal SSD architecture with zero communication and aggregation cost by 21.7$\times$ (15.5$\times$) on average for general matrix-matrix multiplication (GEMM)-intensive (convolution-intensive) ML workloads.
This large performance gap reflects three key limitations of existing network-in-SSD techniques. They (1) still rely on the SSD controller to orchestrate all computations and communication, (2) lack computation-aware communication primitives (e.g., broadcast, distributed synchronization), and (3) lack in-network aggregation capabilities essential for efficient distributed IFP.
 
\textbf{Our goal} is to scale distributed IFP by enabling flash chips to exchange and aggregate partial results directly.

\head{Key Ideas} We propose \textit{\textbf{\namePaper{}}}, a lightweight, decentralized IFP-aware SSD architecture, based on three key ideas.
First, to enable direct chip-to-chip communication,
\namePaper{} builds an IFP network by embedding a lightweight IFP-aware router adjacent to each flash chip and connecting neighboring routers via short, high-speed links. These routers initiate data transfers within the flash array without the controller's orchestration.
\namePaper{} supports multiple network topologies (e.g., ring, mesh, tree),  while preserving compatibility with commodity NAND flash chips.
Second, to coordinate distributed computations efficiently and reduce communication overhead, \namePaper{} introduces lightweight IFP-aware communication primitives (e.g., a broadcast command that starts bit-serial multiplication in all flash chips) for operand distribution, synchronization, and in-network aggregation. 
The SSD controller launches the computation, and routers coordinate local execution and forward partial results directly to the chips that aggregate them.
Third, to reduce data movement, \namePaper{} enables complex arithmetic (e.g., matrix multiplication) within the flash chips and performs simple computations (e.g., comparison, shifting, and masking) in the routers during data transfer.

\head{Case Study} GEMM (e.g.,~\cite{guo2025transitive, qin2020sigma}) and convolution (e.g.,~\cite{Shafiee-isca-2016, lu2021optimizing}) operations are central to machine learning training and inference pipelines (e.g.,~\cite{sandler2018mobilenetv2, tan2019efficientnet, he2016deep, szegedy2016rethinking, ronneberger2015u, jiao2020tinybert, sanh2019distilbert, radford2019language, brown2020language, agarwal2025gpt}). 
These operations repeatedly move large tensors between storage and computation units, making them well-suited for IFP.
When computation spans multiple chips, both operations require aggregation of partial results.  
\namePaper{} reformulates these operations into distributed vector-scalar multiplications by partitioning input tensors across the flash chips and distributing weight elements as scalars for local computation. 
Each flash chip's router (1) forwards partial results directly to a designated flash chip, and (2) supports simple non-linear transformations (i.e., comparison, bit shifts, and masking) during transit.

\head{Evaluation and Results} Using an in-house SSD simulator\footnote{
We develop \namePaper{}'s simulator because existing SSD simulators do not support networked IFP. Our simulator (1) inherits its core SSD model from the state-of-the-art SSD simulator, MQSim~\cite{tavakkol-fast-2018, mqsim-github}, (2) models performance and energy consumption of in-flash computations from prior IFP characterization studies~\cite{park2022flash, chen2024aresflash, kabra2025ciphermatch, kim2025crossbit}, and (3) models router communication, link traversal, and controller interactions based on prior networked SSD techniques~\cite{nadig2023venice, tavakkol2012network, cui2024smartnetssd, kim2022networked}. We calibrate our simulator using real-device characterization studies (e.g.,~\cite{park2022flash, chen2024aresflash}) and open-source infrastructures (e.g.,~\cite{qemu}) (see \S\ref{sec:evaluation_methodology}).}, we compare \namePaper{} with the host CPU, GPU, ISP, IFP on a shared-channel SSD~\cite{chen2024aresflash}, and IFP on a state-of-the-art network-in-SSD architecture~\cite{nadig2023venice}. 
Compared to IFP on a state-of-the-art network-in-SSD architecture~\cite{nadig2023venice}, \namePaper{} improves performance by 2.6$\times$ (2.5$\times$) and reduces energy consumption by 1.3$\times$ (1.7$\times$) on average for five GEMM- (convolution-) intensive workloads.
\namePaper{}'s benefits come with low area and power overheads: each router chip adds printed circuit board area equal to 8\% of a flash die's footprint~\cite{kang-isscc-2019} and consumes 2.2 mW, and each link consumes 1.08 mW during a 4 KB page transfer.

\noindent This work makes the following key contributions.
\squishlist
    \item We identify shared-channel organization and controller-orchestrated execution as two architectural bottlenecks in a modern solid-state drive (SSD) that limit the scalability of in-flash processing (IFP). 
    \item We propose \namePaper{}, the first scalable IFP framework that enables (1) parallel IFP across the flash chips, (2)  direct chip-to-chip communication without the controller's mediation, and (3) in-network aggregation of partial results. 
    \item We show that \namePaper{} accelerates general matrix-matrix multiplication and convolution by reformulating them as distributed vector-scalar operations.
    \item Across ten ML workloads, \namePaper{} provides significant performance and energy efficiency benefits over IFP on a state-of-the-art network-in-SSD architecture.  
\squishend
\section{Background\label{sec:background}}
\subsection{SSD Overview \label{subsec:ssd_overview}}
\begin{figure}[b]
    \centering
    \includegraphics[width=\linewidth]{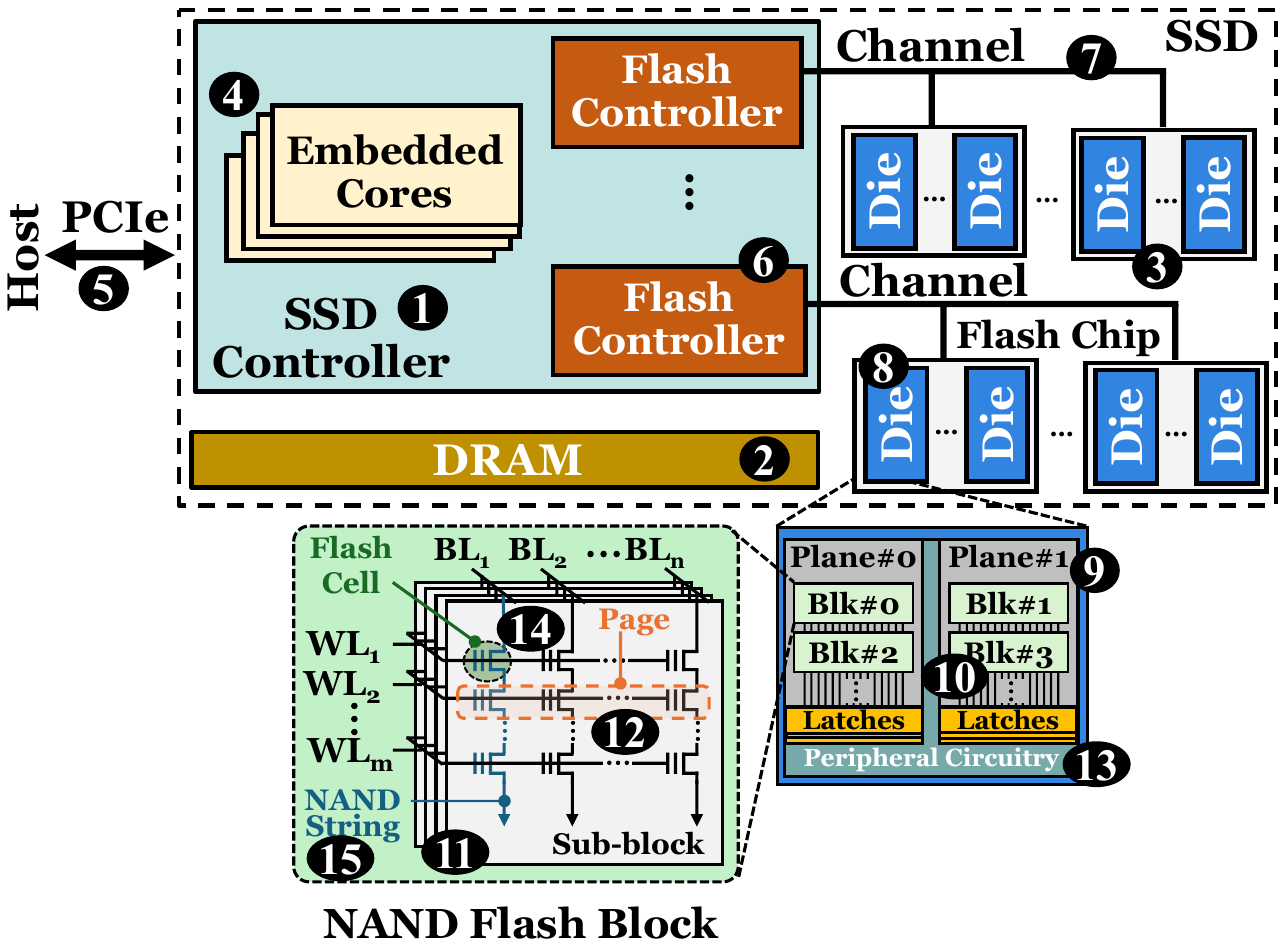}
    % \vspace{-1em}
    \caption{High-level overview of a modern SSD.}
    \label{fig:ssd_architecture}
\end{figure}

Figure~\ref{fig:ssd_architecture} shows a modern NAND flash-based SSD with its three key components: SSD controller~\circled{1}, DRAM~\circled{2}, and NAND flash chips~\circled{3}.

The SSD controller contains multiple general-purpose embedded cores~\circled{4}~\cite{arm-cortexR5,arm-cortexR8}, which run the Flash Translation Layer (FTL) firmware~\cite{gupta2009dftl, tavakkol-fast-2018, lim2010faster, shin2009ftl, zhou2015efficient, sun_leaftl_2023}. 
The FTL manages five key functions: (1) host communication over SATA~\cite{sata} or NVMe~\cite{nvme} (over PCIe~\circled{5}), (2) logical-to-physical (L2P) address translation~\cite{gupta2009dftl, sun_leaftl_2023, huang2015flashmap}, (3) garbage collection~\cite{yang2014garbage, cai-procieee-2017, tavakkol-fast-2018, agrawal2008design, shahidi2016exploring, lee2013preemptible,jung2012taking, choi2018parallelizing,wu2016gcar,cai.bookchapter18.arxiv,cai-hpca-2017}, (4) wear-leveling~\cite{cai-date-2012, cai-dsn-2015, cai-hpca-2015, cai-hpca-2017,cai-iccd-2012, chang2007endurance, huang2017flashblox}, and (5) caching of frequently accessed pages in SSD DRAM.
The SSD controller includes multiple flash controllers (FCs)~\circled{6}~\cite{kim2021decoupled, kim2022networked,wu2012reducing, micheloni-insidenand-2010, microchip-16-channel-controller}.
Each FC connects to a group of flash chips~\cite{cai-procieee-2017, agrawal2008design} through shared channels~\circled{7}.
An FC performs three tasks: (i) command/data transfer~\cite{micheloni-insidenand-2010, onfi-2024}, (ii) data randomization~\cite{cai-procieee-2017,cai-hpca-2017,cai.bookchapter18.arxiv}, and (iii) ECC encoding and decoding~\cite{cai-procieee-2017, cai-hpca-2017, zhao-fast-2013, micheloni-insidenand-2010, tanakamaru-jssc-2013, park-asplos-2021}.
The flash chips on a shared channel can operate concurrently, but only one chip can transfer data over the channel at a time~\cite{nadig2023venice, tavakkol2012network, kim2022networked}.
As a result, this shared-channel organization serializes data transfers between the flash chips and the SSD controller (see \S\ref{subsec:motivation_shared_channel}).

A NAND flash chip~\cite{stoica2019understanding, yucai-thesis} contains 1-4 dies~\circled{8}. 
Each flash die has multiple planes~\circled{9}, and each plane has thousands of blocks~\circleds{10}.
Each block is composed of many sub-blocks~\circleds{11}, each of which has hundreds of pages~\circleds{12}. 
Each flash die includes peripheral circuitry~\circleds{13} with a sensing latch and data latches that temporarily stage data during program and read operations. 
A flash cell~\circleds{14} encodes data using the threshold voltage (high for 0 (programmed), low for 1 (erased)) and acts as an open switch or as a resistor that controls current flow during reads.
A flash cell stores one bit in single-level cell (SLC) mode and three bits in triple-level cell (TLC) mode.
SLC mode provides wider voltage margins between threshold-voltage states than TLC mode, which makes SLC more reliable~\cite{cai-procieee-2017}.
Vertically stacked flash cells (typically 24–176) connect in series to form a NAND string~\circleds{15}, which connects to a bitline (BL). 
The cells at the same vertical position of a sub-block share a wordline (WL). 
A page consists of one bit of each cell on a WL of a sub-block. 

The flash chip reads and programs data at page granularity (e.g., 16 KiB), and senses all BLs of a page in parallel~\cite{cai-procieee-2017, micheloni-insidenand-2010, yucai-thesis}. 
To read a page, the flash chip applies a read-reference voltage to the target WL and a pass voltage to all other WLs of the block, which turns on all non-target cells. 
As a result, each target cell controls the current of its NAND string: a programmed target cell acts as an open switch, and an erased target cell acts as a resistor that conducts current.

\subsection{Storage-Centric Computing \label{subsec:background_ifp}}
Storage-centric computing techniques (e.g.,~\cite{park2016storage, kabra2025ciphermatch, mailthody-micro-2019, kim-fast-2021, kang-msst-2013, torabzadehkashi-pdp-2019, seshadri-osdi-2014, wang-eurosys-2019,acharya-asplos-1998,keeton-sigmod-1998, wang-damon-2016, koo-micro-2017, tiwari-fast-2013, tiwari-hotpower-2012, boboila-msst-2012, bae-cikm-2013,torabzadehkashi-ipdpsw-2018, pei-tos-2019, do-sigmod-2013, kim-infosci-2016, riedel-computer-2001,riedel-vldb-1998, liang-atc-2019,cho-wondp-2013, jun-isca-2015, lee-ieeecal-2020, ajdari-hpca-2019,liang-fpl-2019,jun-hpec-2016, kim-sigops-2020, lee2022smartsage, li2023ecssd, ruan2019insider, wang2016ssd1, li-atc-2021, jeong-tpds-2019, gouk2024dockerssd, ghiasi_megis_2024, ghiasi2022genstore, kang-micro-2021, yavits2021giraf, kim2023optimstore, lim-icce-2021, narasimhamurthy2019sage, jun-isca-2018, fakhry2023review, gu-isca-2016, yang2023lambda, jo2016yoursql, chen2025reis, park2022flash, gao-micro-2021, chen2024aresflash, kim2025crossbit, kang-tc-2021, mahapatra2025rag, lee2017extrav, wang2024beacongnn, yu2024cambricon,jaliminchecs, pan2024instinfer, kang2024isp, soysal2025mars, nadig2026conduit}) move computation closer to data in storage devices. 
We describe two classes of these techniques: in-storage processing (ISP) and in-flash processing (IFP).

ISP techniques~\cite{park2016storage, kabra2025ciphermatch, mailthody-micro-2019, kim-fast-2021, kang-msst-2013, torabzadehkashi-pdp-2019, seshadri-osdi-2014, wang-eurosys-2019,acharya-asplos-1998,keeton-sigmod-1998, wang-damon-2016, koo-micro-2017, tiwari-fast-2013, tiwari-hotpower-2012, boboila-msst-2012, bae-cikm-2013,torabzadehkashi-ipdpsw-2018, pei-tos-2019, do-sigmod-2013, kim-infosci-2016, riedel-computer-2001,riedel-vldb-1998, liang-atc-2019,cho-wondp-2013, jun-isca-2015, lee-ieeecal-2020, ajdari-hpca-2019,liang-fpl-2019,jun-hpec-2016, kim-sigops-2020, lee2022smartsage, li2023ecssd, ruan2019insider, wang2016ssd1, li-atc-2021, jeong-tpds-2019, gouk2024dockerssd, ghiasi_megis_2024, ghiasi2022genstore, kang-micro-2021, yavits2021giraf, kim2023optimstore, lim-icce-2021, narasimhamurthy2019sage, jun-isca-2018, fakhry2023review, gu-isca-2016, yang2023lambda, jo2016yoursql, chen2025reis} integrate computation units (e.g., embedded processors, FPGAs) near the SSD controller to accelerate computations (e.g., filtering, sorting). 
ISP reduces host data transfer, but still requires moving data from flash dies to computation units over bandwidth-limited flash channels, which remain a bottleneck~\cite{park2022flash}.

\begin{figure}[t]
    \centering
    \includegraphics[width=\linewidth]{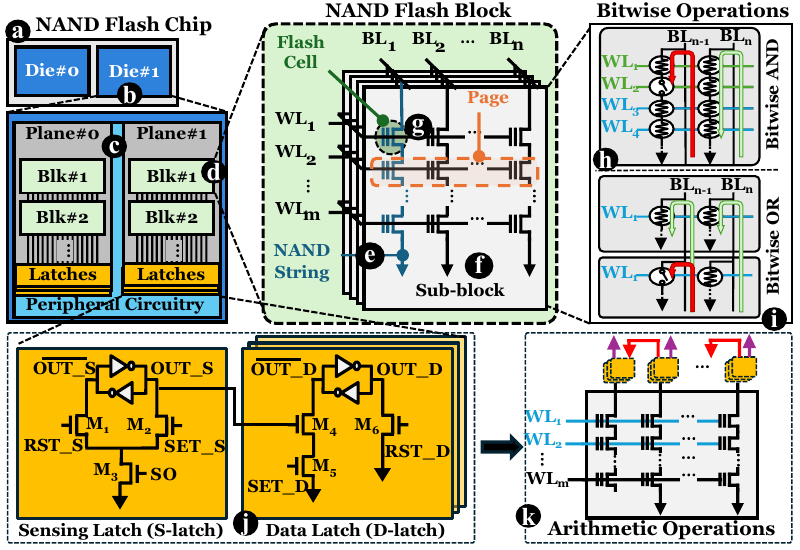}
    \caption{Principles of In-Flash Processing.}
    \label{fig:background_ifp}
\end{figure}

Prior IFP techniques~\cite{park2022flash, gao-micro-2021, chen2024aresflash, kim2025crossbit, kang-tc-2021, chen2024search, wong2024tcam, chen2025reis, jung2026kif} perform computations directly inside NAND flash chips.
They exploit the flash chips' operating principles and peripheral circuitry to perform computations \emph{during a read operation}, as shown in Figure~\ref{fig:background_ifp}. 
IFP techniques exploit the SSD's massive internal parallelism and reduce data transfer to the SSD controller.
Prior IFP techniques support two types of operations: (1) bulk bitwise operations, and (2) arithmetic operations.

\head{IFP: Bulk Bitwise Operations}
Prior IFP works~\cite{park2022flash, gao-micro-2021, chen2024aresflash, kim2025crossbit} perform bulk bitwise operations in two ways. 
First, Flash-Cosmos~\cite{park2022flash} proposes multi-wordline sensing, which senses multiple wordlines at once.
Sensing multiple WLs in the same block computes bitwise AND~\circleds{h} based on the key idea: the cells of a NAND string connect in series, so the string conducts current only if all of its target cells conduct.
Sensing WLs in different blocks computes bitwise OR~\circled{i} based on the key idea: the NAND strings in different blocks connect to the same BL in parallel, so the BL conducts current if any of its strings conducts.
Second, other techniques~\cite{chen2024aresflash,kabra2025ciphermatch,gao-micro-2021, kim2025crossbit} control the sensing latch (S-Latch) and data latches (D-Latches)~\circleds{j} to enable bitwise AND and OR operations. These techniques require reading two operands from the flash array to the peripheral circuitry. 
Commodity SSDs already support bitwise XOR between data latches~\cite{chen2024search,kim-ieeejssc-2012, kim-ieeejssc-2018}, and bitwise NOT using the inverse read operation~\cite{lee-ieeejssc-2002, lee-isscc-2002}. 

\head{IFP: Arithmetic Operations}
Prior works~\cite{chen2024aresflash, kabra2025ciphermatch} support complex operations~\circleds{k} such as bit-serial addition and multiplication. 
These works (1) implement full adder logic using bitwise AND and OR operations, (2) shift a carry bit by adding transmission wires across latches connected to different bitlines, and (3) perform multiplication as repeated shift-and-add operations.
\section{Motivation \label{sec:motivation}}
We examine the limitations of prior shared-channel (\S\ref{subsec:motivation_shared_channel}) and network-in-SSD architectures (\S\ref{subsec:motivation_networked}), and their effectiveness for IFP (\S\ref{subsec:motivation_effectiveness}).

\subsection{Limitations of a Shared-Channel SSD \label{subsec:motivation_shared_channel}}
Modern SSDs connect tens of flash chips to the SSD controller through a small number of shared channels~\cite{nadig2023venice, kim2022networked, cai-procieee-2017, agrawal2008design}. 
In distributed IFP, each flash chip performs computations on its local data and produces partial results, which must leave the flash chip for aggregation.
This organization constrains distributed IFP in two ways.

\head{Bandwidth Bottleneck} 
The flash array provides massive parallelism by performing in-flash computations (during page reads) across many flash chips concurrently.
However, the shared channels expose only a small fraction of the internal bandwidth. 
To demonstrate this bottleneck, we consider an example SSD with eight channels, with each channel connecting eight flash chips. 
Each flash chip contains one die with two planes, a page size of 16 KiB, and an SLC page read latency of 25 $\mu$s~\cite{park2022flash}.
When all planes operate in parallel, the flash array sustains a maximum internal read bandwidth of

\noindent $BW_{Read}$ = \(\frac{64\ \textrm{chips}\ \times\ 1\ \textrm{die}\ \times\ 2\ \textrm{planes}\ \times\ 16\textrm{KiB}}{25\mu \textrm{s}}\) $\approx$ 80 GiB/s

\noindent In contrast, the aggregate bandwidth of the eight shared channels is
$BW_{transfer}$ = 8 $\times$ 1.2 GiB/s = 9.6 GiB/s, assuming 1.2 GiB/s channel bandwidth~\cite{onfi-2024}.
This results in an 8$\times$ lower bandwidth than the internal read bandwidth.
DRAM queueing, command arbitration, and ECC decoding overheads~\cite{nadig2023venice, kim2022networked, tavakkol2012network} further exacerbate this bandwidth bottleneck.
Because \emph{only} one flash chip can transfer data over a channel at a time~\cite{nadig2023venice, kim2022networked, tavakkol2012network}, the shared channels serialize data transfers, which increases the overall read latency. 
As more flash chips participate in IFP, the shared-channel architecture becomes increasingly saturated, fundamentally limiting IFP's scalability.

\begin{figure}[b]
    \centering
    \includegraphics[width=0.9\linewidth]{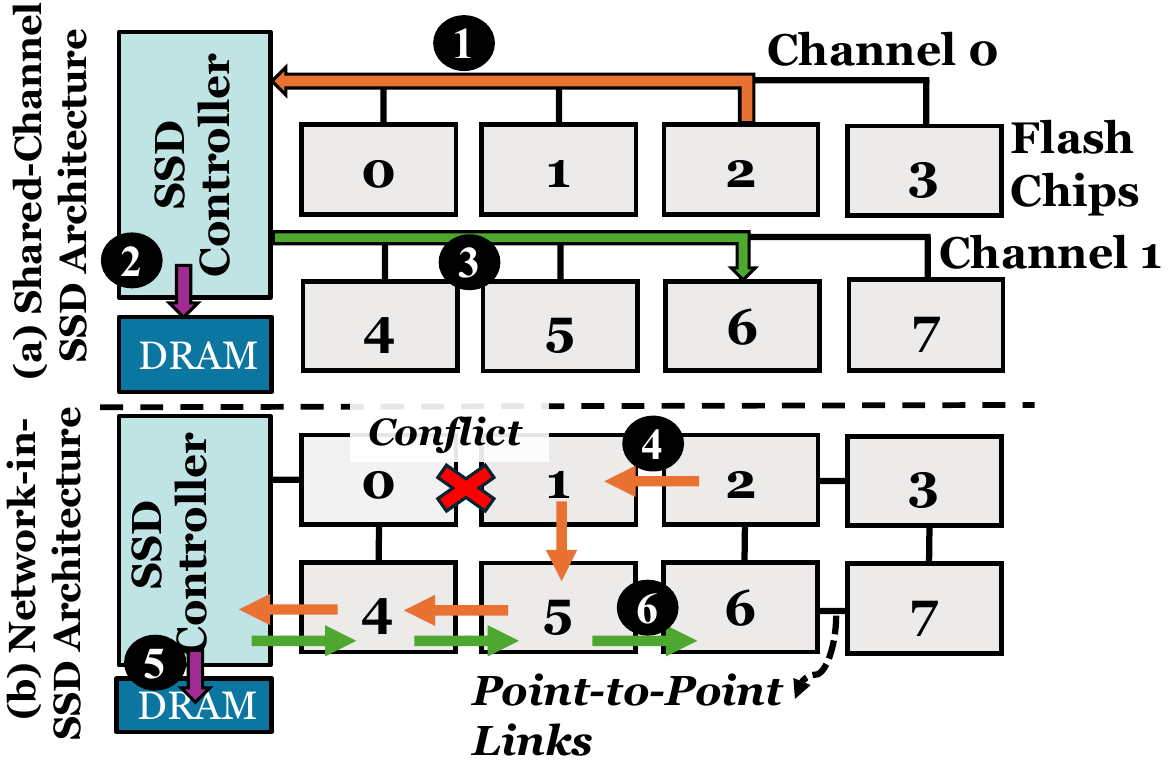}
    \caption{Data movement in (a) a shared-channel SSD architecture, and (b) a network-in-SSD architecture.}
    \label{fig:motivation_comparison}
\end{figure}

\head{Controller-Orchestrated Execution}
The SSD controller orchestrates all computation, data transfers, and aggregation~\cite{nadig2023venice, kim2022networked, nadig2026conduit, kim2021decoupled}, which prevents each flash chip from directly communicating with its neighboring flash chips.
As \fig~\ref{fig:motivation_comparison}(a) shows, a partial result generated in flash chip 2 traverses a shared channel to the SSD controller \circled{1}. 
The SSD controller stages the partial result temporarily in the SSD DRAM \circled{2}, and then forwards it over a shared channel to the destination flash chip \circled{3}, which aggregates it.
This multi-stage data path requires multiple channel transfers and DRAM accesses for every partial result, which substantially increases latency and energy consumption during IFP aggregation. 
As distributed workloads become larger, the controller increasingly becomes the communication bottleneck, which fundamentally limits IFP's scalability. 

\subsection{Limitations of Network-in-SSD Architectures \label{subsec:motivation_networked}}

To overcome the limitations of a shared-channel SSD, prior network-in-SSD architectures (e.g.,~\cite{nadig2023venice,kim2022networked,tavakkol2012network,cui2024smartnetssd}) embed an interconnection network of flash chips within the SSD.
These architectures replace shared channels with short point-to-point links (\fig~\ref{fig:motivation_comparison}(b)).
The links provide multiple paths between the SSD controller and each flash chip (i.e., path diversity).
Path diversity reduces serialized data transfers and improves storage I/O performance. 
For example, Venice~\cite{nadig2023venice} reserves a conflict-free path in the network before it initiates a data transfer. 
However, these architectures \emph{do not} efficiently support distributed IFP for two key reasons.

First, they still rely on the SSD controller to orchestrate every computation and data transfer. 
As shown in \fig~\ref{fig:motivation_comparison}(b), the SSD controller transfers a partial result from the source flash chip over a conflict-free path \circled{4}.
The SSD controller stages the partial result temporarily in the SSD DRAM \circled{5}, and forwards it to the destination chip \circled{6}.
A conflict-free path reduces data serialization and latency. 
However, the flash chip still cannot communicate directly with its neighboring flash chips, which limits IFP scalability.

Second, these architectures lack computation-aware communication primitives.
Data-intensive workloads (e.g., machine learning) frequently require one-to-many and many-to-one communication patterns~\cite{krishna2011towards, graham2016sharp, sapio2021scaling, lao2021atp, kwon2018maeri}. 
For example, GEMM and convolution operations repeatedly distribute common operands to multiple flash chips and aggregate partial results produced across those chips. 
Current network-in-SSD architectures expose point-to-point communication, so a broadcast requires repeated unicast transfers.
These repeated transfers increase network traffic, create communication hotspots, and limit IFP scalability.

We conclude that prior network-in-SSD architectures optimize storage I/O rather than distributed IFP. 
Hence, naively combining them with IFP-capable chips is insufficient.

\subsection{Effectiveness of Prior SSD Architectures \label{subsec:motivation_effectiveness}}
\head{Methodology}
We study the effectiveness of prior SSD architectures on ten end-to-end ML workloads.
We use an in-house SSD simulator that builds on the state-of-the-art SSD simulator, MQSim ~\cite{tavakkol-fast-2018, mqsim-github} (see \S\ref{sec:evaluation_methodology} for our methodology). 
We compare three architectures.
First, \emph{\baseline{}} enables state-of-the-art IFP primitives~\cite{chen2024aresflash, park2022flash, kim2025crossbit} on flash chips in a shared-channel SSD.
Second, \emph{\venice{}} augments the state-of-the-art network-in-SSD architecture, Venice~\cite{nadig2023venice}, with IFP-capable flash chips~\cite{chen2024aresflash, park2022flash, kim2025crossbit}.
Third, \emph{\ideal{}} uses the same network as \venice{}, but assumes zero-latency chip-to-chip communication and controller-free aggregation. \ideal{} serves as an upper bound for distributed IFP. 
These architectures have the same computation latency, but differ in communication latency.

\begin{figure}[t]
    \centering
    \includegraphics[width=\linewidth]{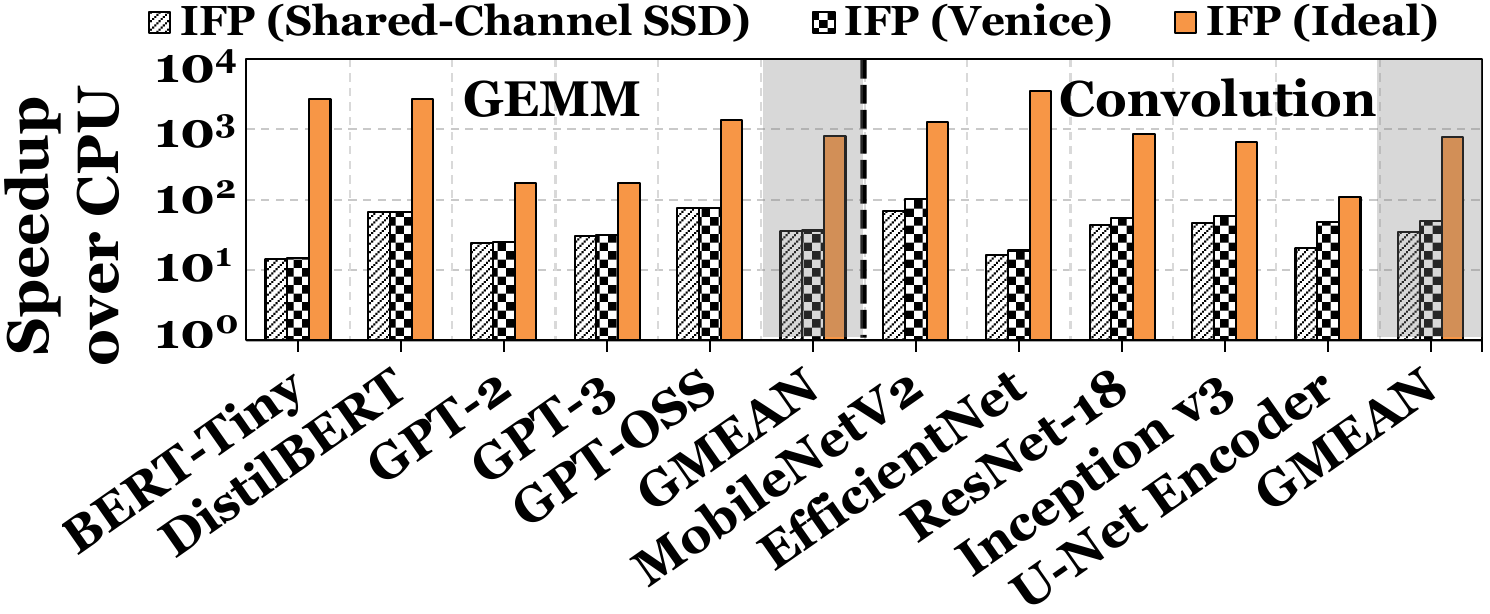}
    \caption{Speedup of \baseline{}, \venice{} and \ideal{} over CPU. Higher is better.}
    \label{fig:motivation}
\end{figure}

\head{Performance Results}
\fig~\ref{fig:motivation} shows the speedup of the three architectures over the host CPU on five GEMM-intensive and five convolution-intensive ML workloads.  

We make two observations.
First, \venice{} improves performance by 1.1$\times$ (1.5$\times$) over \baseline{} for GEMM- (convolution-) intensive workloads.
\venice{}'s benefits come from improved path diversity and conflict-free transfers of intermediate results during aggregation.
Second, \venice{} falls short of \ideal{} by 21.7$\times$ (15.5$\times$) on average for GEMM- (convolution-) intensive workloads.
Because all three architectures have the same computation latency, this gap comes entirely from communication and aggregation.
This large performance gap stems from \venice{}'s controller-orchestrated data transfer and communication overheads.

We conclude that replacing an SSD's shared-channel organization with Venice's network-in-SSD architecture provides only modest performance benefits for distributed IFP.
  
\textbf{Our goal} is to scale distributed IFP by enabling flash chips to exchange and aggregate partial results directly.
\section{\namePaper{} \label{sec:mechanism}}
\head{Overview} We propose \namePaper{}, a lightweight and scalable IFP-aware SSD architecture that enables flash chips to exchange and aggregate partial results directly.
\namePaper{} is based on three key ideas. 
First, to enable direct chip-to-chip communication, \namePaper{} builds an IFP network by embedding a lightweight IFP-aware router adjacent to each flash chip and connecting neighboring routers via short, high-speed links. 
These routers initiate data transfers within the flash array without the SSD controller's orchestration. 
Second, to coordinate distributed computations efficiently and reduce communication overhead, \namePaper{} introduces lightweight IFP-aware communication primitives (e.g., a broadcast command that initiates IFP across all flash chips) for operand distribution, synchronization, and in-network aggregation. 
Third, to reduce data movement, \namePaper{} enables complex arithmetic (e.g., matrix multiplication) within the flash chips, and performs simple computations (e.g., comparison, shifting and masking) in the routers during data transfer. 
We describe Lattice's network (\S\ref{subsec:lattice_network}), node (\S\ref{subsec:lattice_node}), router chip (\S\ref{subsec:lattice_router}), packet format (\S\ref{subsec:lattice_packetformat}), and flow control (\S\ref{subsec:lattice_flowcontrol}).

\subsection{Network for Distributed IFP \label{subsec:lattice_network}}
\begin{figure}[t]
    \centering
    \includegraphics[width=\linewidth]{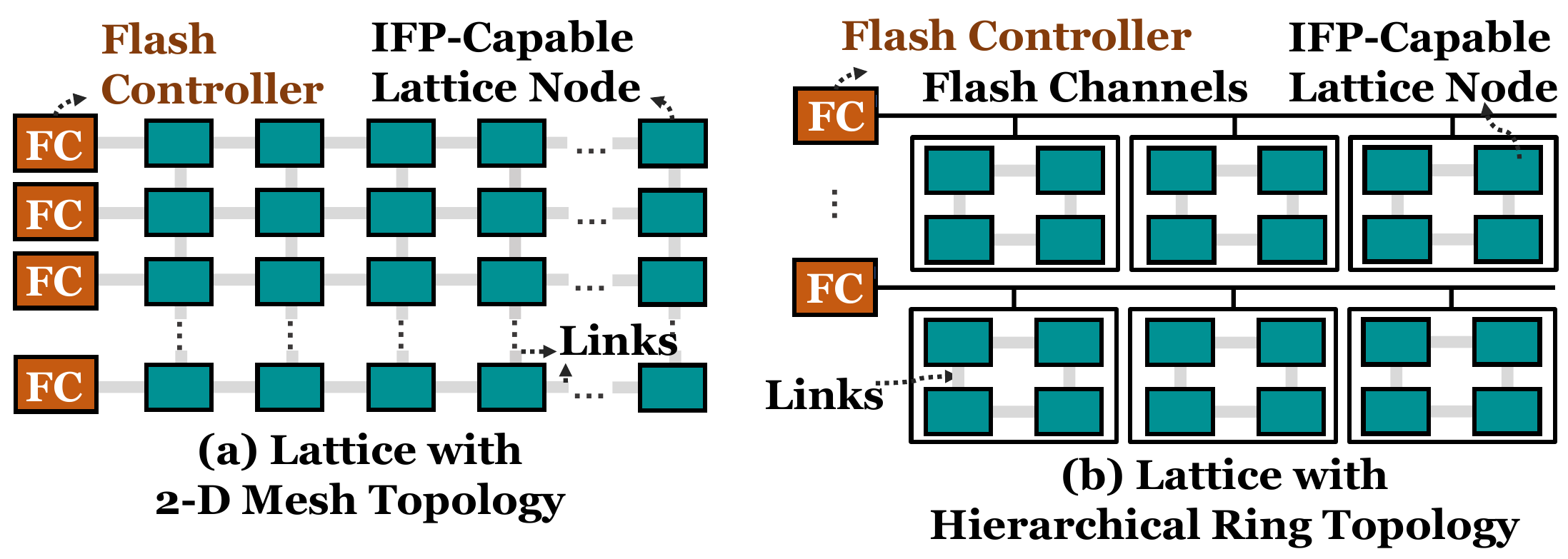}
    \caption{\namePaper{} with example network topologies.}
    \label{fig:mechanism_overview}
\end{figure}

\namePaper{} transforms an SSD from a set of isolated flash chips into a computation network of IFP-capable nodes.
\namePaper{} supports multiple network topologies (e.g., ring, mesh, tree) based on application requirements.
\fig~\ref{fig:mechanism_overview} shows two example topologies.
In a 2-D mesh topology (\fig~\ref{fig:mechanism_overview}(a)), \namePaper{} connects each network node with up to four neighbors (north, east, south, west). 
In a hierarchical ring topology (\fig~\ref{fig:mechanism_overview}(b)), \namePaper{} connects network nodes in concentric rings or chains.
We discuss \namePaper{}'s applicability to other topologies in \S\ref{sec:discussion}.

\subsection{\namePaper{} Node \label{subsec:lattice_node}}
\begin{figure}[b]
    \centering
    \includegraphics[width=\linewidth]{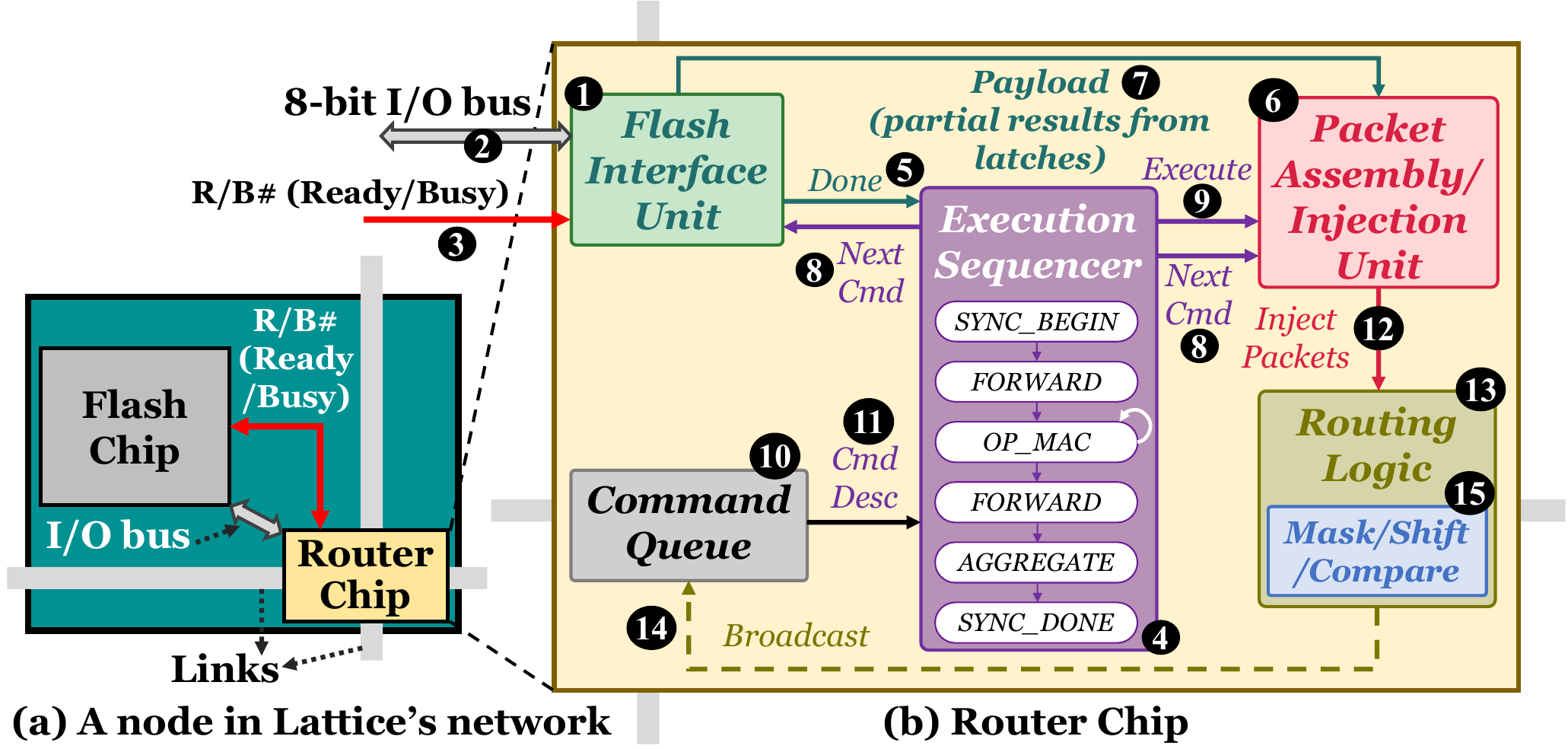}
    % \vspace{-1em}
    \caption{Structure of (a) \namePaper{}'s node and (b) router chip.}
    \label{fig:mechanism_router}
\end{figure}

\fig~\ref{fig:mechanism_router}(a) shows the structure of \namePaper{}'s node.
Each node includes two key components.
First, an IFP-capable flash chip, that performs bulk bitwise and arithmetic operations (e.g., bulk bitwise operations, addition, multiplication, and shift operations). 
The flash chip's design builds on prior in-flash computation techniques~\cite{park2022flash, gao-micro-2021, chen2024aresflash, kabra2025ciphermatch}.
Second, a lightweight, IFP-aware router chip placed adjacent to the flash chip.
\namePaper{} separates the router chip from the flash chip in the node to avoid modifications to the flash chip's package. 
The router connects to the flash chip through the flash chip's standard 8-bit I/O bus and its ready/busy (R/B\#) pin~\cite{onfi-2024}, and to neighboring nodes' routers through short bidirectional point-to-point links.
\namePaper{} preserves the passive execution model of commodity NAND flash chips: the router performs all communication and packet assembly, and the flash chip performs computations only in response to the router's commands.

\subsection{Router Chip \label{subsec:lattice_router}}
The router chip (\fig~\ref{fig:mechanism_router}(b)) is an IFP-aware logic block placed adjacent to each flash chip and connected to neighboring routers via short bidirectional point-to-point links.
The router has four key elements: (1) four input/output ports (north, south, east, west for 2-D mesh topology), (2) one local injection/ejection port connected to the flash chip, (3) small per-port input buffers (12 bytes each), and (4) a lightweight 8-bit crossbar datapath that connects input ports to output ports and matches the standard NAND flash interface width.
\namePaper{}'s router has six components.

\head{1. Flash Interface Unit (FIU)}
The FIU~\circled{1} interfaces with the flash chip through the standard 8-bit NAND I/O interface~\circled{2} and R/B\# (ready/busy) pin~\circled{3} that already exist in commodity flash chips.
The FIU performs three key tasks: it (1) delivers the flash commands (e.g., bit-serial multiply-and-accumulate and addition) issued by the execution sequencer~\circled{4} to the flash chip, (2) monitors the flash chip's R/B\# pin and issues a \texttt{Done} event~\circled{5} to the execution sequencer, and (3) reads the partial results out of the flash chip's data latches and streams them to the packet assembly unit~\circled{6} as payload~\circled{7}.

\head{2. Execution Sequencer}
The execution sequencer~\circled{4} controls the flash chip's computation.
It implements a per-tile finite-state machine (FSM) that tracks the flash chip's current computation phase.
The FSM has six phases: SYNC\_BEGIN $\rightarrow$ FORWARD (receive scalar operand) $\rightarrow$ OP\_MAC (loop over scalar operands) $\rightarrow$ FORWARD (send partial result) $\rightarrow$ AGGREGATE $\rightarrow$ SYNC\_DONE (see \S\ref{subsec:lattice_packetformat}).
The FSM advances to the next state when it receives the \texttt{Done} signal~\circled{5} from the FIU.
The execution sequencer produces two outputs: it sends (1) the next computation or synchronization command~\circled{8} to the FIU or the packet assembly unit, respectively, and (2) an execute signal~\circled{9} to the packet assembly unit. 
This unit requires only a 3-bit state register, two counters, and a comparator per in-flight tile.

\head{3. Command Queue}
The command queue~\circled{10} is a small SRAM-based queue that (i) buffers controller-issued commands and synchronization events (e.g., SYNC\_BEGIN), and (ii) enforces ordering dependencies between computation and communication.
\namePaper{} forwards the command descriptors (e.g., Destination, TileID, Operation)~\circled{11} to the execution sequencer for further processing.

\head{4. Packet Assembly/Injection Unit}
The packet assembly and injection unit~\circled{6} assembles packets and injects them into the network.
When the sequencer issues the \texttt{Execute} command, the packet assembly and injection unit~\circled{6} constructs the packet header while the FIU streams the partial result from the flash chip.
This unit (1) encapsulates the payload with routing metadata and operation information, (2) segments the stream into flits\footnote{A flit is the smallest unit of flow control~\cite{dally2004principles}.} over the shared 8-bit datapath without intermediate buffering, and (3) injects packets~\circled{12} into the network via the routing logic.

\head{5. Routing Logic}
The routing logic~\circled{13} forwards each packet towards its destination.
It uses a deterministic routing algorithm (e.g., dimension-ordered X-Y routing~\cite{dally1987deadlock, dally2004principles}) that minimizes hardware complexity, guarantees transfer latency, and avoids deadlocks in \namePaper{}'s network. 
The routing logic supports three communication primitives:
(1) \textit{Unicast}, for point-to-point operand transfer and partial-result aggregation.
(2) \textit{Multicast}, for distributing operands and partial results to multiple flash chips.
(3)~\textit{Broadcast}, for propagating global commands (e.g., IFP\_START) and distributing scalar weights to all chips.
Each router replicates broadcast packets~\circled{14} hop by hop, which eliminates per-chip commands and reduces channel saturation.

The routing logic is a shallow pipeline with three stages: (1) packet-header decoding (see \S\ref{subsec:lattice_packetformat}), (2) output-port allocation, and (3) crossbar traversal. 
Each stage completes within one cycle at a 1.2 GHz router frequency. 
A single-hop traversal incurs three router cycles plus serialized link-transfer latency, depending on the payload length and link bandwidth. 

\head{6. In-Network Operation Units}
Each router integrates small combinational units~\circled{15} that perform simple element-wise transformations during packet traversal. 
These units perform four operations: (1) masking, for predicate filtering and conditional forwarding, (2) simple comparison, for selection and reduction primitives, (3) shifting, for operand alignment and block-based number scaling, and (4) zero-value suppression, for sparse-data propagation and traffic reduction.
The router applies these transformations in flight, without additional buffers or routing-pipeline stalls. 

\subsection{Packet Format \label{subsec:lattice_packetformat}}
A \namePaper{} packet consists of a 96-bit header (see Table~\ref{tab:packet}) and a flit-based payload~\cite{dally2004principles} that carries operands and partial results.
The header has ten key fields.

\begin{table}[b]
\caption{\namePaper{}'s Packet Header}
\label{tab:packet}
\centering
\renewcommand{\arraystretch}{1.2}
\setlength{\tabcolsep}{2pt}
\resizebox{\linewidth}{!}{%
\begin{tabular}{|c||c|c|}
\hline
\textbf{Field} & \textbf{Size (in bits)} & \textbf{Description} \\ 
\hline
\hline
Command Type & 4 & {\begin{tabular}[c]{@{}c@{}} Operations \\  (e.g., IFP\_START, SYNC\_BEGIN) \end{tabular}}  \\
\hline
Opcode & 4 & {\begin{tabular}[c]{@{}c@{}} In-flash operation to execute \\ (e.g., OP\_LOAD\_PAGE, OP\_MAC) \end{tabular}}  \\
\hline
Flags & 4 & Packet handling (e.g., RETRY, ACK, DROP) \\
\hline
{\begin{tabular}[c]{@{}c@{}} Tile ID 
\end{tabular}} & 12 & {\begin{tabular}[c]{@{}c@{}} Physical block that stores the operands \\ (e.g., tensor tile, graph block) 
\end{tabular}} \\
\hline
{\begin{tabular}[c]{@{}c@{}} Page ID 
\end{tabular}} & 12 & {\begin{tabular}[c]{@{}c@{}} Wordline number of operand 
\end{tabular}} \\
\hline
Source ID & 8 & Source node ID \\
\hline
Destination ID & 8 & Destination node ID \\
\hline
Hop Count & 8 & Remaining hops before packet termination \\
\hline
Payload Length & 16 & Number of payload bytes, excluding header \\
\hline
Metadata & 16 & Application-specific metadata \\
\hline
Reserved Bits & 4 & Reserved for future use \\
\hline
\end{tabular}
 }
\end{table}

\head{Command Type} 
\namePaper{} defines six key commands that orchestrate IFP execution:
(1) \texttt{SYNC\_BEGIN} (broadcast) instructs all nodes to read operands in flash chips and prepare flash chips' latches for IFP. 
(2) \texttt{IFP\_START} (broadcast) triggers bit-serial computation (e.g., add, multiply, shift) in flash chips.
(3) \texttt{FORWARD} (unicast) transfers operands to a flash chip from the controller or another flash chip.
(4) \texttt{SYNC\_REDUCE} instructs routers to apply in-network data reductions (see \S\ref{subsec:lattice_router}).
(5) \texttt{AGGREGATE} signals the flash chip of the designated aggregation node to combine partial results using in-flash addition. 
(6) \texttt{SYNC\_DONE} indicates completion of aggregation. 

\head{Opcode} The opcode specifies the in-flash operation to execute. 
Key opcodes include: 
(1) \texttt{OP\_LOAD\_PAGE} reads a page into the sensing latch,
(2) \texttt{OP\_SEND\_OPERAND} stores an operand distributed by the SSD controller in the flash chip's data latches,
(3) \texttt{OP\_MAC} performs a bit-serial multiply-and-accumulate operation, 
(4) \texttt{OP\_ADD} performs bit-serial addition of two operands in the data latches~\cite{kim2025crossbit, chen2024aresflash, kabra2025ciphermatch}, 
(5) \texttt{OP\_SHIFT} shifts operands by $n$ bits~\cite{chen2024aresflash}, 
(6) \texttt{OP\_SEND\_PARTIAL} forwards the partial result to the designated aggregation node, and 
(7) \texttt{OP\_REDUCE\_ACCUM} aggregates partial results in the destination flash chip's latches using bit-serial addition.

\head{Flags} The flags provide fine-grained control over packet handling and forwarding (e.g., acknowledgment (ACK) for control packets, retry (RETRY), dropping packets (DROP)). 

\head{Tile ID} The tile ID identifies the physical block address where the operands are stored. 

\head{Page ID} The page ID specifies the physical page address within the tile, enabling precise page-level operand access.

\head{Source ID and Destination ID} These fields specify the source and target flash chips. Destination ID 0xFF denotes broadcast. 

\head{Hop Count} The hop count tracks the remaining routing hops before the packet must be terminated, to ensure livelock-free traversal. 

\head{Payload Length} The payload length specifies the number of payload bytes. 

\head{Metadata} The metadata bits store application-specific metadata (e.g., convolution offsets), which keeps the routers application-agnostic.

\subsection{Flow Control \label{subsec:lattice_flowcontrol}} 
\namePaper{} uses credit-based flow control  (e.g.,~\cite{ray2012scalable, kung1995credit}) to regulate packet traversal across the network. 
Each output port maintains a small credit counter that tracks buffer availability in the next-hop router.
The output port stalls transmission when credits are unavailable.
This approach has three benefits: it (1) avoids large FIFO buffers, (2) provides predictable latency, and (3) prevents congestion.

When multiple packets contend for the same output port, the router resolves contention using a round-robin arbitration policy to ensure fairness across concurrent traffic flows. 
The router transfers packets at flit granularity, enabling flit pipelining and efficient serialization of partial-result vectors across multi-hop paths.
For control packets (e.g., SYNC\_BEGIN, IFP\_START), routers send ACK/NACK responses to the controller to confirm network-wide delivery.
\section{Case Study: Machine Learning Workloads \label{sec:casestudy}}
Modern ML models, from convolutional neural networks (CNNs) (e.g.,~\cite{sandler2018mobilenetv2, tan2019efficientnet, he2016deep, szegedy2016rethinking, ronneberger2015u}) to transformer-based large language models (LLMs) (e.g.,~\cite{jiao2020tinybert, sanh2019distilbert, radford2019language, brown2020language, agarwal2025gpt, vaswani2017attention, devlin2019bert}), are dominated by 
general matrix-matrix multiplication (GEMM) (e.g.,~\cite{guo2025transitive, qin2020sigma, goto2008anatomy}) and convolution (e.g.,~\cite{Shafiee-isca-2016, lu2021optimizing, chetlur2014cudnn}) operations. 
These operations repeatedly move large multi-dimensional tensors across the memory hierarchy, and they rely on massively parallel, reduction-heavy computations.
These characteristics make them well-suited for \namePaper{}'s IFP substrate.

\head{Topology Choice} We adopt a 2-D mesh topology because GEMM and convolution partition computations into rows, columns, or spatial tiles that primarily communicate with neighboring tiles. These communication patterns naturally map to the horizontal and vertical links of a 2-D mesh. 

\begin{figure*}[h]
    \centering
    \includegraphics[width=0.95\textwidth]{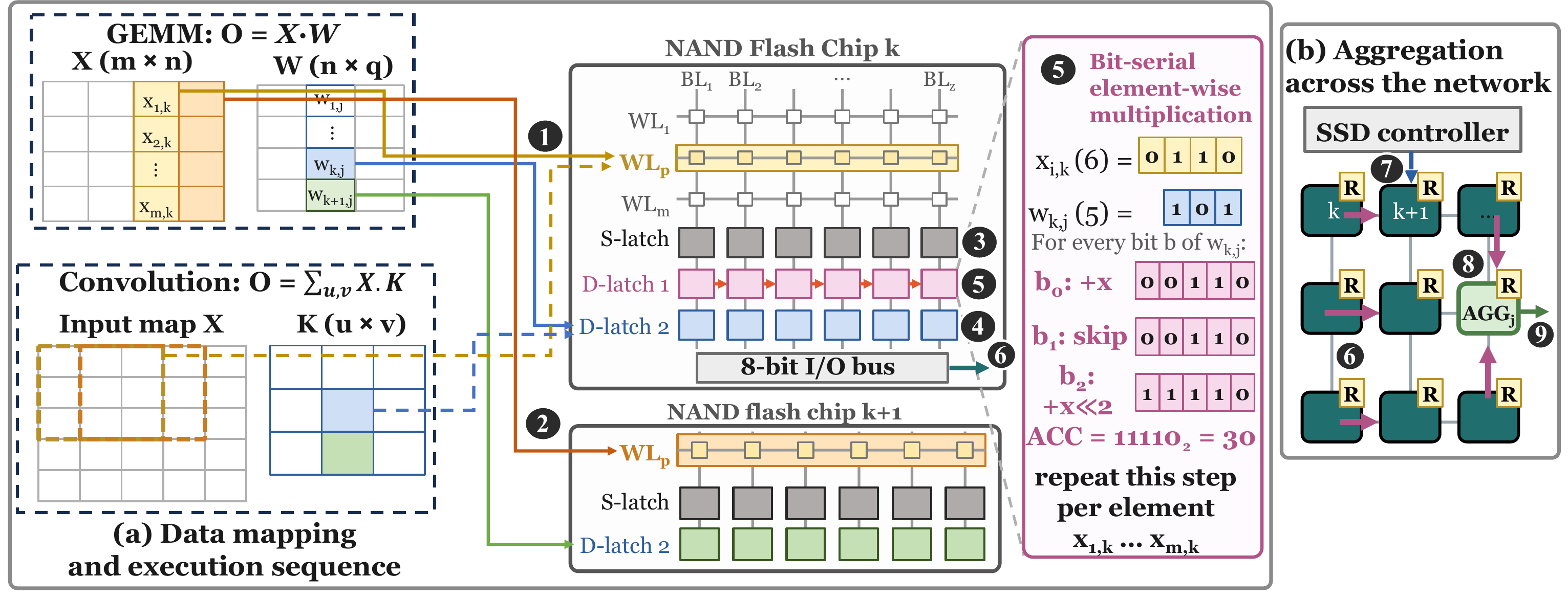}
    % \vspace{-1em}
    \caption{GEMM and Convolution: Data mapping and execution sequence (a) and aggregation across the network (b).}
    \label{fig:casestudy_gemm_convolution}
\end{figure*}

\subsection{GEMM and Convolution\label{subsec:gemm_convolution}}
GEMM computes $\mathbf{O} = \mathbf{X} \mathbf{W} + \mathbf{B}$, where each element of $\mathbf{XW}$ is the dot product of a row of the input matrix $\mathbf{X}$ and a column of the weight matrix $\mathbf{W}$.
Convolution computes each output element $\mathbf{O(i,j)}$ by sliding a filter with different weights (the convolution kernel $\mathbf{K}$) across the input matrix ($\mathbf{X}$).
$O(i,j) = \sum_{u,v} X(i+u, j+v) \cdot K(u,v)$.
Despite their different mathematical formulations, both repeatedly perform (1) multiply-and-accumulate (MAC) operations in which an input vector is scaled by a weight or kernel vector, and (2) accumulation of partial products. 
Existing IFP techniques~\cite{chen2024aresflash, kim2025crossbit} support \emph{only} vector–scalar operations, making conventional dot-product execution difficult to map directly onto flash arrays.
\namePaper{} addresses this limitation by distributing the input across flash chips and reformulating GEMM and convolution as parallel vector-scalar multiplications followed by in-flash accumulation. 
We describe the execution flow for both operations in six steps.

\head{Partitioning Inputs Across Chips}
Instead of row-major placement (each matrix row occupies a flash page), \namePaper{} distributes input matrix columns in GEMM, or spatial tiles (i.e., column vectors) of the input feature map in convolution, across flash chips (see \S\ref{subsec:system_integration} for data placement and distribution strategy).
Each flash chip stores one column vector of $\mathbf{X}$ per page (\circled{1} and \circled{2} in Fig.~\ref{fig:casestudy_gemm_convolution}(a)), so that the flash chip can read the column elements in parallel along the NAND bitlines.
This mapping aligns in-flash vector–scalar multiplication with bitline-level parallelism and allows each flash chip to operate directly on its locally stored data without inter-chip transfers of input operands.
The SSD controller initiates GEMM and convolution with a broadcast command, \texttt{[CommandType: SYNC\_BEGIN, Opcode: OP\_LOAD\_PAGE, TileID: t, PageID: p]}. 
Upon receiving this, each router sends commands to its flash chip to (1) read the local column vector of (\texttt{TileID t}, \texttt{PageID p}) into the sensing latch \circled{3} and (2) initialize data latches to store the computation result or partial-result accumulation.

\head{Distribution of Scalar Operands}
\namePaper{} distributes scalar operands (i.e., elements of the GEMM weight matrix $\mathbf{W}$, or of the convolution kernel $\mathbf{K}$) across its nodes (see \S\ref{subsec:system_integration}) using: \texttt{[CommandType: FORWARD, Opcode: OP\_SEND\_OPERAND, Source: FC ID, Destination: Dst. Node ID, Payload: scalar $w_{k,j}$, TileID: t]}.
\namePaper{} stores scalars in flash chips' data latches \circled{4} to avoid program operations and preserve SSD lifetime.

\head{In-Flash Vector-Scalar Multiplication}
After the input vectors and scalar operands are in the flash chips' latches, \namePaper{} initiates vector-scalar multiplication across all chips in parallel with a single broadcast command, \texttt{[CommandType: IFP\_START, Opcode: OP\_MAC, TileID: t]}, which replaces per-chip commands from the SSD controller. 
Each chip multiplies its local vector by the received scalar using bit-serial shift-and-add operations~\cite{chen2024aresflash} (\circled{5}).

During multiplication, \namePaper{} iterates over the bits of the scalar in the data latches. 
For every set bit $b_i$ of $w_{k,j}$, the flash chip adds a copy of the input vector (in the sensing latch \circled{3}), left-shifted by $i$ bit positions (i.e., $x{\ll}i$), to the data latch (D-latch 1 \circled{5}), which accumulates the result. 
The flash chip skips zero-valued bits.
For example (\circled{5} in Fig.~\ref{fig:casestudy_gemm_convolution}(a)), to multiply $x_{i,k} = 6$ ($0110_2$) by $w_{k,j} = 5$ ($101_2$), the flash chip adds $x_{i,k}$ for $b_0$, skips $b_1$, and adds $x_{i,k} \ll 2$ for $b_2$, which yields 30 ($11110_2$). 
Because each bitline holds one element of the stored vector, this shift-and-add sequence executes concurrently for all elements $x_{1,k},\dots,x_{m,k}$ across the flash array's bitlines.
This sequence produces a partial result: one term of the dot product in GEMM, or the partial output $P^{(u,v)}(i,j) = X(i{+}u, j{+}v) \cdot K(u,v)$ for kernel offset $(u,v)$ in convolution. 
\namePaper{} repeats this step for each scalar operand. 
Executing these operations in parallel exploits the flash array's full internal bandwidth and bitline-level MAC capability.

\head{Forwarding Partial Results}
After each multiplication, the router forwards the partial-result tile over chip-to-chip links to a designated aggregation node (\circled{6} in Fig.~\ref{fig:casestudy_gemm_convolution}(a) and (b)), via a unicast command: \texttt{[CommandType: FORWARD, Opcode: OP\_SEND\_PARTIAL, Source: Src. Node ID, Destination: Dst. Node ID, Payload: $P_i^{(j)}$ or $P^{(u,v)}(i,j)$, TileID: t]}.
This step bypasses the SSD controller and repeats for each scalar operand.

\head{Aggregation at the Destination Flash Chip}
The aggregation node's flash chip receives the partial results and performs in-flash reduction in its data latches using bit-serial addition~\cite{kabra2025ciphermatch, chen2024aresflash, kim2025crossbit}. 
\namePaper{} initiates this step with \texttt{[CommandType: AGGREGATE, Opcode: OP\_REDUCE\_ACCUM, Source: FC ID, Destination: Dst. Node ID, TileID: t]} (\circled{7}), which instructs the flash chip to accumulate all partial results tagged with \texttt{TileID} $= t$ ~(\circled{8}) (see Fig.~\ref{fig:casestudy_gemm_convolution}(b)). 
For GEMM, this reduction completes the dot product. 
For convolution, the aggregation node accumulates incoming partial maps across kernel offsets until it processes the full convolution window to generate the final output tile $O_i$.

\namePaper{} supports strided convolution~\cite{he2016deep, krizhevsky2012imagenet, simonyan2014very}. 
For stride $\mathbf{s}$: (1) each node computes partial results, and (2) the aggregation node retains only every $\mathbf{s}$-th position. 
Routers optionally suppress (i.e., mask and drop using their mask/shift unit) non-stride elements during forwarding, reducing inter-chip communication.

\head{Completion and Handoff to the Next Layer}
After aggregation completes, the aggregation node's router sends a completion notification, \texttt{[CommandType: SYNC\_DONE, Opcode: OP\_NOP, TileID: t]}, back to the controller. 
This notification indicates that the aggregated output is ready for the next stage (\circled{9}) (e.g., transferring the output to the host, forwarding it to the next layer's GEMM or convolution, or writing it to the flash page).

\subsection{Block Floating-Point Number Support \label{subsec:mechanism_bfp}} 
Prior IFP techniques have limited accuracy because they support only fixed-point or integer arithmetic.
Conventional floating-point representation requires storing and aligning per-value exponents, which is incompatible with the NAND flash chip's bit-serial computation model.
To overcome this limitation, \namePaper{} adopts the block floating-point (BFP) representation~\cite{drumond2018hybridbf, rouhani2020msfp, rouhani2023microscaling, ieee754-2019}, which balances accuracy and hardware simplicity. 
BFP has low storage overhead and computation complexity because it performs exponent alignment only once per block instead of per value. 

In BFP, a group of values shares a common exponent, while each value stores only its mantissa. 
Each value is represented as $a = m_a \cdot 2^{e}$, where $m_a$ is the individual mantissa and $e$ is the shared exponent.
\namePaper{} stores mantissas directly in the flash array as integers, and the exponent in each flash page's out-of-band (OOB) area~\cite{qin2021atomicwrites}.

A MAC operation in BFP expands to: 
$o = x \cdot w + y = m_x \cdot 2^{e_x} \cdot m_w \cdot 2^{e_w} + m_y \cdot 2^{e_y} = m_x \cdot m_w \cdot 2^{e_x + e_w} + m_y \cdot 2^{e_y}$.
To add these terms, \namePaper{} aligns the exponents. 
If $e_x + e_w > e_y$, the mantissa of $y$ is right-shifted by $s=e_x + e_w - e_y$ before accumulation.
If $e_x + e_w < e_y$, the mantissa of $w$ (or $x$) is shifted before multiplication.
\namePaper{} performs exponent alignment inside the network in three steps.
First, each router detects exponent mismatches in the packet header metadata.
Second, the router right-shifts the mantissas with its in-network operation units.
Third, the aggregation node aggregates the aligned mantissas without additional exponent logic.

\subsection{Support for Nonlinear Operations}
ML models include lightweight nonlinear layers (e.g., ReLU~\cite{nair2010relu}, max pooling) between GEMM and convolution stages. 
Executing these layers on the host requires transferring full activation maps through the memory hierarchy, incurring significant data movement. 
\namePaper{} reduces this overhead by executing these layers inside the SSD. 

The router integrates in-network operation units that operate on data as it traverses \namePaper{}'s network.
These units perform three operations.
First, MSB-based masking for thresholding operations (e.g., ReLU), which checks each mantissa's sign bit and replaces negative values with zeros. 
Second, compare-and-forward logic for reduction-style operations (e.g., max pooling), which repeatedly uses lightweight pairwise comparisons to identify the larger value until a single pooled value remains.
Third, shift units for exponent alignment during BFP normalization, which shift mantissas based on the shared exponent metadata.

\subsection{System Integration \label{subsec:system_integration}}
We describe five aspects of integrating \namePaper{} into an SSD.

\head{Operand Placement and Distribution Strategy} \namePaper{} stores input activations or feature-map tiles in flash chips, and distributes weights across flash chips because different flash chips use the weight matrix simultaneously during GEMM and convolution. 
This strategy has three benefits: it (1) avoids duplicate weight copies, (2) reduces program operations and flash wear, because weights reside only in flash chips' latches during computation (see \S\ref{subsec:gemm_convolution}), and (3) minimizes communication overhead and aligns with modern accelerator dataflow strategies~\cite{jouppi2017datacenter, chen2016eyeriss}.

\namePaper{} does not require fixed operand placement. 
While \namePaper{} primarily distributes weights, it can also distribute activations if weights reside in flash.  
If both weights and activations reside in the flash chip, \namePaper{} reads them into the flash chip's latches for computation without modifying the communication substrate.

\head{Router Design Considerations}
\namePaper{} preserves compatibility with commodity SSDs by adding routers as lightweight companion logic blocks adjacent to each flash chip without modifying flash packages. 
Alternatively, SSD designers can integrate routers within the flash package or as chiplet-style companion dies connected via short interconnects.
The routers operate in a synchronous timing domain aligned with the flash-interface clock, avoiding incompatibility between flash devices and routing logic.

\head{System Interfaces} 
\namePaper{} exposes five new interfaces to the host system to perform in-flash GEMM and convolution: (1) \textbf{\namePaper{}\_GEMM (TileID, X, W, Dimensions, BFP\_Exponent)}, (2) \textbf{\namePaper{}\_Conv (TileID, X, K, Stride, BFP\_Exponent)}, (3) \textbf{\namePaper{}\_ReLU (TileID)}, (4) \textbf{\namePaper{}\_MaxPool (TileID, Window, Stride)}, and (5) \textbf{ \namePaper{}\_ReadOutput (TileID)}.

Each call specifies input tile IDs, weight or kernel pointers, stride or pooling parameters, and optional BFP metadata.
The SSD controller translates these calls into controller commands (e.g., SYNC\_BEGIN, IFP\_START) (see \S\ref{sec:mechanism}, Table~\ref{tab:packet}). 

\head{Data Reliability}
The dominant reliability challenge in distributed IFP is the NAND flash array, where threshold-voltage variation affects in-flash computation. 
To address this challenge, \namePaper{} adopts Enhanced SLC-Mode Programming (ESP)~\cite{park2022flash}, which widens voltage margins to improve read reliability without controller-side ECC decoding. 
The tradeoff is reduced storage capacity, as blocks operating in SLC mode store fewer bits than blocks in TLC mode. 
To address the capacity overhead, \namePaper{} partitions the SSD into two regions: (1) a \emph{\namePaper{} region} that operates in SLC mode and stores data participating in IFP, and (2) a \emph{conventional region} that continues to operate in TLC mode for conventional storage.  

After computation, \namePaper{} (1) transfers intermediate results over short point-to-point links, and (2) temporarily buffers them in the router's SRAM buffers. 
\namePaper{}'s short links are inherently more reliable~\cite{dally1998digital} than shared channels because they exhibit (1) significantly reduced parasitic capacitance, (2) lower crosstalk, and (3) stable voltage margins. 
We calibrate the timing parameters and perform impedance matching to improve reliability. 
The aggregation router acknowledges the packets, and the source router retransmits packets at fixed intervals until it receives an acknowledgment. 
\namePaper{} follows the same reliability assumptions as the SRAM components and network buffers widely used in modern processors and on-chip networks. 
We leave stronger error detection and correction mechanisms to future work.

\head{FTL Integration} 
\namePaper{} does not modify address translation, garbage collection, or wear-leveling, but operates as a complementary execution substrate. 
During execution, the FTL allocates and pins pages for input tiles and output buffers to prevent relocation or garbage collection.

\subsection{Overhead Analysis \label{subsec:results_overhead}}
\head{Area Overhead} 
\namePaper{}'s routers and links introduce no modifications to NAND flash chips, because each router is implemented as a separate chip adjacent to a flash chip. 
Hence, the routers and links add area only to the SSD's printed circuit board (PCB).

We estimate the router area by synthesizing its hardware description language (HDL) model in a 45 $nm$ technology node~\cite{nangate_45nm, nangate_opencelllib, stine2007freepdk}. 
Based on the router's design (see \S\ref{subsec:lattice_router}), the resulting router area is 0.0057 $mm^2$, which is negligible compared to the footprint of modern flash dies ($\simeq$100 $mm^2$ ~\cite{kang-isscc-2019}).
I/O pads, not logic, dominate each router chip's PCB area.
Each router exposes 40 I/O pins (8-bit bidirectional links and two control pins on each side), which occupy 8 $mm^2$ of PCB area.
Each router chip thus adds PCB area equal to 8\% of a flash die area.

We estimate \namePaper{}'s link area using ORION 3.0~\cite{kahng2015orion3}. 
An 8$\times$8 2-D mesh (see Table~\ref{tab:eval_config}) requires 112 network links (excluding injection/ejection links, which are the same as channel connectors).
Each link occupies $\approx$ $0.04\times$ the shared channel area, so the 112 links occupy 44\% less total interconnect area than eight shared channels.
Compared to a shared bus, network links are 8$\times$ shorter and narrower, as they require lower pitch sizes.

\namePaper{} uses the in-flash arithmetic operations proposed by prior work (e.g.,~\cite{chen2024aresflash}), which add transistors and transmission wires to the peripheral circuitry of flash chips.
These additions incur an area overhead of 0.84\% of the flash chip area.

\noindent\textbf{Power Overhead.} 
We estimate the router's power consumption by synthesizing the design for the 45 $nm$ technology node~\cite{nangate_45nm, nangate_opencelllib, stine2007freepdk}.
Each router consumes 2.2 mW.
Using ORION 3.0~\cite{kahng2015orion3}, we estimate that each network link consumes 1.08 mW~\cite{nadig2023venice} during a 4KB NAND flash page transfer.
Compared to a shared channel, each network link consumes 90\% less power because the link (1) is shorter and narrower than a shared bus, and (2) has only two drivers compared to several drivers in a shared bus (e.g., eight in an $8\times$8 flash array configuration).
\section{Evaluation Methodology \label{sec:evaluation_methodology}} 
\subsection{Simulator Overview}
To evaluate \namePaper{} and the baselines, we build an in-house event-driven SSD simulator on top of MQSim~\cite{tavakkol-fast-2018, mqsim-github}, a state-of-the-art SSD simulator that is validated against real SSDs.
We extend MQSim to model (1) in-flash computation and (2) \namePaper{}'s network topology and direct chip-to-chip transfers.
Table~\ref{tab:eval_config} describes our evaluated configurations.

\head{SSD Modeling} \namePaper{}'s simulator builds on MQSim's SSD model, which includes (1) cycle-level modeling of I/O operations (program, erase, read), 
(2) L2P address translation, (3) transaction scheduling across channels and dies, (4) NVMe interface for host-device communication, (5) resource contention at channel and die levels, and (6) garbage collection and wear-leveling.
We implement a demand-based L2P mapping cache similar to DFTL~\cite{gupta2009dftl}.

\head{Network Modeling}
Each flash chip connects to a lightweight router (see \S\ref{subsec:lattice_router}) that connects to neighboring routers via short point-to-point links in an 8×8 2-D mesh topology.
We implement a cycle-level network model that captures three effects: (1) link traversal latency, based on the configured link bandwidth (1.2 GiB/s) and payload size, (2) serialization, by transferring packets at flit granularity, and (3) contention at router input ports and links.
The model supports (i) unicast, multicast, and broadcast communication, (ii) forwarding of partial results across routers, and (iii) in-network operations (e.g., masking, comparison, and shift).
We configure our network model consistently with prior network-in-SSD architectures~\cite{nadig2023venice, kim2022networked, tavakkol2012network, cui2024smartnetssd}, which report similar link bandwidths, flow-control techniques, and contention models.

\noindent \textbf{Performance Modeling.}
We model the latency of in-flash computation with four components: (1) command transfer latency, (2) flash sensing latency (i.e., $T_{Read}$), (3) latency of data transfer from page buffer to flash controller (i.e., $T_{DMA}$), and (4) latency of arithmetic operations (e.g., addition, multiplication) in the peripheral circuitry (see \S\ref{sec:background}). 
We validate this model against real-device characterization and results from prior work (e.g., \cite{chen2024aresflash, kabra2025ciphermatch, park2022flash, gao-micro-2021, nadig2026conduit}).
We model computation on SSD controller cores with QEMU~\cite{qemu, bellard2005qemu} configured for ARM Cortex-R8~\cite{arm-cortexR8}.

\noindent\noindent\textbf{Energy Modeling.} 
We model the energy consumption of (1) flash chips based on Samsung 980 Pro's~\cite{samsung-980pro, park2022flash} power values, (2) computations on the controller cores, using ARM Cortex-R8 power models~\cite{arm-cortexR8}, and (3) DRAM, using DDR4 power values~\cite{ghose2019demystifying}.

\head{Metrics} We report (1) speedup over the host CPU, in terms of overall execution time of the workloads, and (2) energy consumption normalized to that of the host CPU.

\begin{table}[t]
  \centering
  \footnotesize
  \caption{Evaluated Configurations.}
  \label{tab:eval_config}
  \renewcommand{\arraystretch}{1.1}
  \resizebox{\columnwidth}{!}{
  \begin{tabular}{|c||l|}
  \hline
  
  \multirow{5}{*}{
  \centering 
  \shortstack{\textbf{Host CPU} \\ \textbf{(Real System)}}}
  & Intel(R) Xeon(R) Gold 5515+~\cite{intel-xeon-gold-5515} \\ \cline{2-2} 
  & x86-64~\cite{guide2016intel}, 8~cores, out-of-order, 3.2 GHz  \\ \cline{2-2}
  & \emph{L1 Data + Inst. Private Cache:} 48 KiB, 64B line \\ \cline{2-2}
  & \emph{L2 Private Cache:} 2 MiB, 64B line \\ \cline{2-2}
  & \emph{L3 Shared Cache:} 22.5 MiB, 64B line \\ 
  
  \hline

  \multirow{4}{*}{
  \centering 
  \shortstack{\textbf{Host GPU} \\ \textbf{(Real System)}}}
  & NVIDIA A100 GPU~\cite{a100, choquette2020nvidia}, \\
  & NVIDIA Ampere Architecture~\cite{ampere} (7nm) \\ \cline{2-2} 
  & 108 Streaming Multiprocessors, 1.4 GHz base clock;  \\ \cline{2-2}
  & \emph{L2 Cache:} 40 MiB; Main Memory: 40 GiB HBM2~\cite{lee2016simultaneous, lee2014hbm} \\ 
  
  \hline
  
  \multirow{4}{*}{\centering \shortstack{\textbf{Host Main Memory} \\ \textbf{(Real System)}}}
  & 32 GiB DDR4-2400, 4~channels, 1 rank, 16 banks; \\ \cline{2-2} 
  & Peak throughput: 19.2 GiB/s;  \\ \cline{2-2} 
  & \textbf{Latency:} \texttt{T\textsubscript{bbop}}: 49 ns;  \textbf{Energy:} \texttt{E\textsubscript{bbop}}: 0.864 nJ; \\ 
  & where $bbop$ is bulk bitwise operation \\ \cline{2-2} 
  
  \hline
  
  % \multirow{2}{*}{\centering \shortstack{\textbf{Storage (SSD)}}} & Samsung 980 Pro PCIe 4.0 NVMe SSD 2 TB~\cite{samsung-980pro} \\ \cline{2-2} 
  % & External PCIe Bandwidth: 8GB/s \\ \cline{2-2}
  % \hline

  \multirow{11}{*}{
\shortstack{\textbf{SSD~\cite{samsung-980pro}} \\ \\ \textbf{(Simulated)}}}
  & 48-WL-layer 3D TLC NAND flash-based SSD; 2 TiB\\ \cline{2-2}
  & \textbf{External Bandwidth}: PCIe 4.0, 8 GiB/s\\ \cline{2-2} 
  & \textbf{NAND Config}: 8 channels; 8 dies/channel; 2 planes/die;\\
  & 2,048 blocks/plane; 192 (4$\times$48) WLs/block; 4 KiB/page\\ \cline{2-2} 
  & \textbf{Flash Channel Bandwidth}: 1.2 GiB/s\\ \cline{2-2} 
  & \textbf{Latency}: T\textsubscript{read} (SLC mode): 25 $\mu$s~\cite{park2022flash}; \\
  & T\textsubscript{prog} (SLC mode): 400  $\mu$s~\cite{park2022flash}; T\textsubscript{bers} : 3500 $\mu$s~\cite{park2022flash}\\
  & T\textsubscript{AND/OR}: 20 ns~\cite{gao-micro-2021}; T\textsubscript{latchtransfer}: 20 ns~\cite{gao-micro-2021}; \\  
  &  T\textsubscript{XOR}: 30 ns~\cite{park2022flash}; T\textsubscript{DMA}: 13.3 $\mu$s (for 16 KiB);\\\cline{2-2} 
  & \textbf{Energy}: \texttt{E\textsubscript{read}} (SLC mode): 20.5$\mu$J/channel ~\cite{park2022flash}; \\ &
  E\textsubscript{AND/OR}: 10nJ/KiB~\cite{gao-micro-2021}; E\textsubscript{latchtransfer}: 10nJ/KiB~\cite{gao-micro-2021}; 
  \\ & E\textsubscript{XOR}: 20nJ/KiB~\cite{park2022flash}; E\textsubscript{DMA}: 7.656$\mu$J/channel; \\

  \hline

  \multirow{2}{*}{\centering \shortstack{\textbf{SSD DRAM} \\ \textbf{(Simulated)}}} 
  & 2 GiB LPDDR4-1866 DRAM cache;\\
  & 1 channel, 1 rank, 8 banks\\ \cline{2-2} 
  % & \textbf{Latency:} T\textsubscript{bbop}: 49 ns;  \textbf{Energy:} E\textsubscript{bbop}: 0.864 nJ; \\ 
  % & where $bbop$ is bulk bitwise operation \\ 
  \hline 
  
  \textbf{SSD Controller Cores}
  & ARM Cortex-R8 series @1.5GHz; 5 Cores~\cite{arm-cortexR8} \\
  
  \hline
  \multirow{4}{*}{
    \begin{tabular}{c}
    \textbf{\namePaper{} Network Design}
    \end{tabular}
    }
    & \textbf{Topology.} 8$\times$8 2-D mesh, 8-bit 1.2 GiB/s link bandwidth \\
    & \textbf{Router.} 1.2 GHz frequency \\
    & \textbf{Routing Algorithm.} Deterministic \\
    & \textbf{Switching.} Circuit switching~\cite{dally2004principles} \\
    \hline
  \end{tabular}
  }
\end{table}

\begin{table*}[b]
\centering
\scriptsize
\caption{Evaluated end-to-end workloads. \textbf{Layers} counts the transformer
blocks (GEMM) or convolution/fully-connected layers (Conv) executed in flash.}
\label{tab:workloads}
\renewcommand{\arraystretch}{1}
\begin{tabular}{|c||c|c|c|c|c|c|c|} \hline
\textbf{Model} & \textbf{Use Case} & \textbf{Type} & \textbf{Layers} & \textbf{Layer Dimensions} & \textbf{Precision} & \textbf{Sparsity} & \textbf{Batch} \\
\hline
\hline

BERT-Tiny~\cite{turc2019wellread}  & Sentiment Analysis  & GEMM & 2 blocks & $128 \times 128$ -- $128 \times 512$   & 4-bit & 25\% & 64 \\ \hline

DistilBERT~\cite{sanh2019distilbert} & Question Answering & GEMM & 6 blocks & $768 \times 768$ -- $768 \times 3072$  & 4-bit & 40\% & 64   \\ \hline

GPT-2~\cite{radford2019language}  & Text Generation  & GEMM & 12 blocks & $768 \times 768$ -- $768 \times 3072$  & 8-bit & 50\% & 16  \\ \hline

GPT-3 Medium~\cite{brown2020language}  & Text Generation & GEMM & 24 blocks & $1024 \times 1024$ -- $1024 \times 4096$ & 8-bit & 50\% & 16  \\ \hline

GPT-OSS~\cite{agarwal2025gpt} & LLM Training & GEMM & 24 blocks & $1536 \times 1536$ -- $1536 \times 6144$ & 8-bit & 30\% & 128 \\ \hline \hline

MobileNetV2~\cite{sandler2018mobilenetv2} & Image Classification & Conv & 52 Conv, 1 FC & kernel: $1{\times}1$--$3{\times}3$, 16--1280 filters & 4-bit & 35\% & 64 \\ \hline

EfficientNet~\cite{tan2019efficientnet} & Image Classification  & Conv & 49 Conv, 1 FC & kernel: $1{\times}1$--$5{\times}5$, 16--1280 filters & 4-bit & 38\% & 128 \\ \hline

ResNet-18~\cite{he2016deep} & Image Classification & Conv & 20 Conv, 1 FC & kernel: $1{\times}1$--$7{\times}7$, 64--512 filters & 8-bit & 45\% & 128 \\ \hline

Inception v3~\cite{szegedy2016rethinking} & Image Classification & Conv & 94 Conv, 1 FC & kernel: $1{\times}1$--$7{\times}7$, 32--448 filters & 8-bit & 46\% & 64 \\ \hline

U-Net Encoder~\cite{ronneberger2015u} & Semantic Segmentation & Conv & 8 Conv & kernel: $3{\times}3$, 64--512 filters & 8-bit & 55\% & 16
\\ \hline

\end{tabular}
\end{table*}

\subsection{Evaluated Approaches}
\head{Host CPU and GPU} The host CPU and GPU (see Table~\ref{tab:eval_config} for configurations) execute all computations by reading operands from the SSD via the NVMe interface.
We construct strong host baselines by (1) executing each application natively with architecture-specific optimizations (e.g., CUDA for GPU), (2) modeling the SSD-host transfers at the theoretical maximum PCIe bandwidth, (3) modeling the host DRAM latency and bandwidth (see Table~\ref{tab:eval_config}), and (4) accounting for CPU/GPU memory caching, allowing reused operands to avoid redundant SSD and PCIe transfers.

\head{In-Storage Processing (ISP)} \isp{} performs all computations on ARM Cortex-R8 cores~\cite{arm-cortexR8} in the SSD controller, emulated using QEMU~\cite{qemu}.  
ISP transfers operands from flash chips to the controller over the shared flash channels, and intermediate results are staged in the SSD DRAM. 
This baseline also represents a near-flash accelerator, which remains constrained by flash-channel bandwidth. 

\head{IFP (Shared-Channel SSD)} 
We implement this baseline based on a shared-channel SSD that builds on state-of-the-art IFP techniques~\cite{park2022flash, gao-micro-2021, chen2024aresflash, kabra2025ciphermatch}, which support bitwise and arithmetic operations (see \S\ref{subsec:background_ifp}). 
Since prior IFP works do not natively support GEMM and convolution, we implement them using the primitives proposed in these works.

\head{IFP (Venice)}
We augment the state-of-the-art network-in-SSD architecture, Venice~\cite{nadig2023venice}, with IFP-capable flash chips.
Venice replaces the shared-channel organization with a network of flash chips connected via short point-to-point links, providing path diversity and conflict-free data transfers. 
It uses the same in-flash computation primitives as \baseline{}, and Venice's routers perform the same non-linear computations{} as \namePaper{}'s routers. 
As in \baseline{}, the SSD controller in \venice{} orchestrates all operand transfers, synchronization, and aggregation.

\head{\ideal{}}
This policy serves as an upper bound for the performance of distributed IFP. 
It uses the same network architecture as \venice{} and in-flash computation primitives as \baseline{}, while assuming zero-latency chip-to-chip communication and controller-free aggregation.

\subsection{Evaluated Workloads}
We evaluate \namePaper{} on ten end-to-end ML workloads: five GEMM-intensive transformer models and five convolution-intensive CNNs. 
Table~\ref{tab:workloads} summarizes their key characteristics. 
For each model, \namePaper{} executes all GEMM and convolution operations in transformer blocks, or all convolution layers and fully-connected layers, within the SSD.

The five GEMM-intensive workloads span a wide range of model depths, matrix dimensions, and sparsity levels.
They range from BERT-Tiny~\cite{turc2019wellread}, with two transformer blocks and GEMMs of up to 128×512, to GPT-OSS~\cite{agarwal2025gpt}, with 24 blocks and GEMMs of up to 1536×6144.
DistilBERT~\cite{sanh2019distilbert} (six blocks), GPT-2~\cite{radford2019language} (12 blocks), and GPT-3 Medium~\cite{brown2020language} (24 blocks) cover the sizes in between. 
These workloads evaluate \namePaper{}'s parallel vector-scalar multiplication and multi-hop aggregation.

The five convolution-intensive workloads capture diverse spatial and structural characteristics. 
MobileNetV2~\cite{sandler2018mobilenetv2} and EfficientNet~\cite{tan2019efficientnet} consist of lightweight, depthwise-style, low-channel convolution operations common in edge inference. 
ResNet-18~\cite{he2016deep} uses standard 3×3 convolution with stride-2 downsampling.
Inception-v3~\cite{szegedy2016rethinking} has multi-branch computation with significant concurrency and multicast operations. 
U-Net’s~\cite{ronneberger2015u} encoder layer features a large 128×128 activation map with high sparsity, which tests \namePaper{}’s in-network reduction capabilities.

The host CPU performs the computations that the flash chips and routers do not support (e.g., batch normalization~\cite{ioffe2015batchnorm}).
We include their execution time and the associated data transfers in all reported results.
We perform these computations on the host CPU because they account for a small fraction of each model's computation compared to GEMM and convolution, so the host CPU does not become the performance bottleneck.
\section{Results \label{sec:results}}

\subsection{Performance Analysis \label{subsec: results_performance}}

\begin{figure}[t]
\centering
\includegraphics[width=\linewidth]{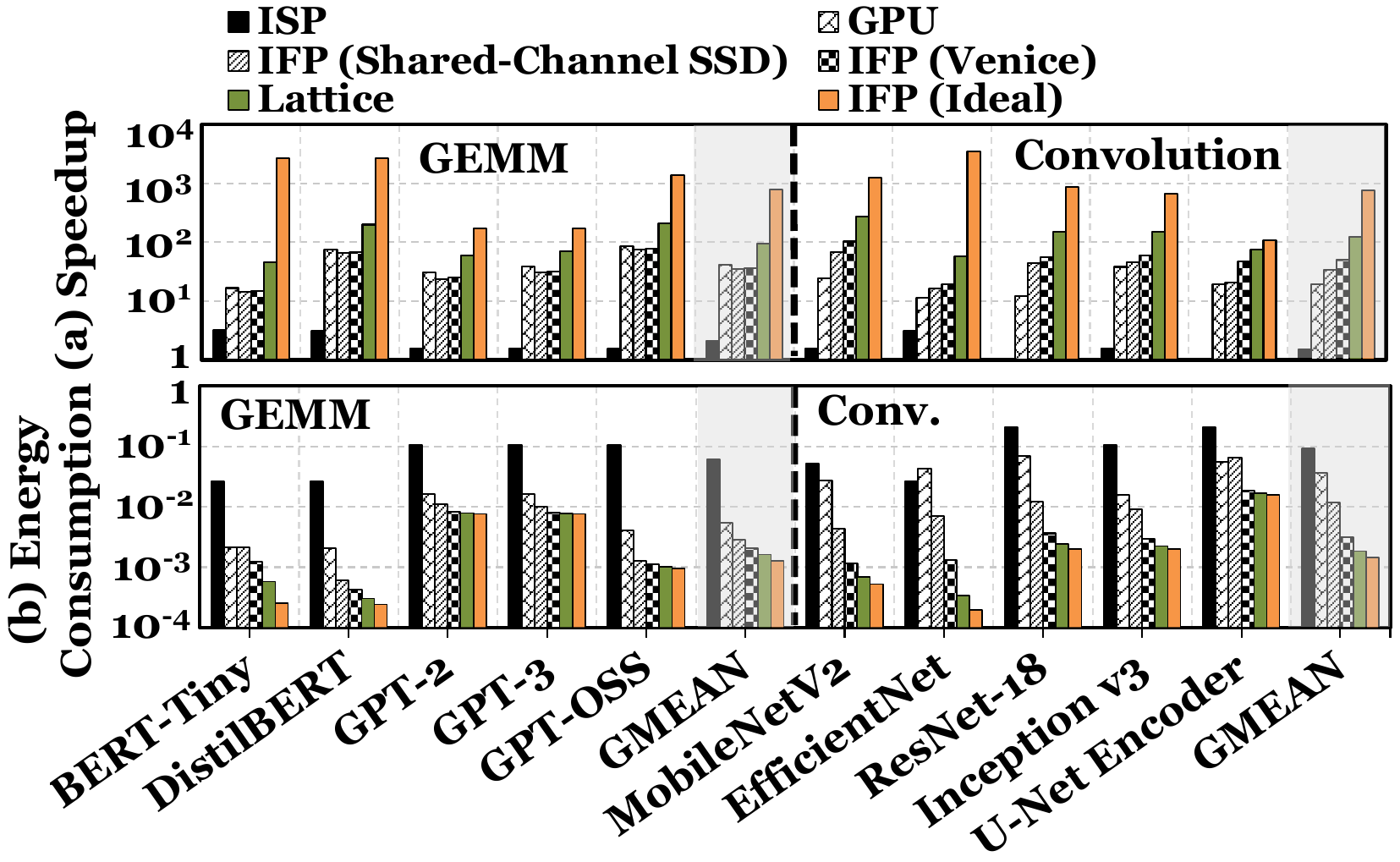}
\caption{Speedup over \cpu{} (a) and energy consumption (b) of \namePaper{} and baselines normalized to those of \cpu{}. Speedup: Higher is better; Energy consumption: Lower is better.}
\label{fig:Speedup}
\end{figure}

\fig~\ref{fig:Speedup}(a) shows the speedup of \namePaper{} and the baselines over \cpu{} for ten ML workloads. We make four key observations.

First, \namePaper{} consistently outperforms all baselines across all workloads. 
On GEMM- (convolution-) intensive workloads, \namePaper{} achieves an average speedup of 95.7$\times$ (122.8$\times$) over CPU, 46.2$\times$ (81.7$\times$) over ISP, 2.3$\times$ (6.4$\times$) over GPU, 2.7$\times$ (3.6$\times$) over \baseline{}, and 2.6$\times$ (2.5$\times$) over \venice{}.
\namePaper{}'s benefits come from its decentralized communication architecture and in-network aggregation.

Second, \namePaper{}'s performance is 8.2$\times$ (6.3$\times$) lower than that of \ideal{} on GEMM- (convolution-) intensive workloads. This performance gap is $\sim$2.5$\times$ narrower than the 21.7$\times$ (15.5$\times$) performance gap between \ideal{} and \venice{}.

Third, on convolution-intensive workloads, \venice{} outperforms CPU by 49.9$\times$, GPU by 2.6$\times$, ISP by 33.2$\times$, and \baseline{} by 1.5$\times$ because (i) convolution maps efficiently to bit-serial computation, and (ii) \venice{} provides path diversity and conflict-free transfers of intermediate results.   
However, \venice{} relies on the SSD controller for repeated synchronization and aggregation of intermediate results, which limits its performance. 
\namePaper{} overcomes this limitation with direct chip-to-chip transfer and in-network aggregation.

Fourth, on GEMM workloads, the GPU outperforms CPU by 42.0$\times$, ISP by 20.2$\times$, \baseline{} by 1.2$\times$, and \venice{} by 1.2$\times$, benefiting from its optimized tensor cores and high compute throughput.
However, PCIe transfers limit the GPU performance. 
\namePaper{} reduces these transfers by executing GEMM inside the SSD and efficiently pipelining vector-scalar multiplications across the network. 
As a result, \namePaper{} outperforms the GPU even on large transformer layers.

We conclude that \namePaper{} significantly improves the performance of distributed IFP by enabling direct chip-to-chip communication and in-network aggregation.

\subsection{Energy Analysis \label{subsec: results_energy}}
\fig~\ref{fig:Speedup}(b) shows the energy consumption of \namePaper{} and baselines over CPU. 
We make five key observations.

First, on all workloads, \namePaper{} has the lowest energy consumption among all evaluated systems except \ideal{}. 
On GEMM- (convolution-) intensive workloads, \namePaper{} reduces energy consumption by 617.5$\times$ (538.3$\times$) over CPU, 37.5$\times$ (49.1$\times$) over ISP, 3.4$\times$ (20.0$\times$) over GPU, 1.8$\times$ (6.3$\times$) over \baseline{}, and 1.3$\times$ (1.7$\times$) over \venice{}.
\namePaper{}'s lower energy consumption comes from its reduced data movement, minimal controller involvement, and in-network aggregation.

Second, \namePaper{} consumes 1.6$\times$ (2.3$\times$) more energy than \ideal{} on GEMM- (convolution-) intensive workloads because \ideal{} assumes zero communication cost.

Third, \venice{} is the most energy-efficient prior approach for both GEMM- and convolution-intensive workloads, but it still performs repeated controller-centric aggregation through the SSD DRAM. 
\namePaper{} avoids this overhead via direct chip-to-chip aggregation and in-network reductions. 

Fourth, in GEMM workloads, the GPU consumes less energy than the CPU and ISP, but it incurs energy overheads from PCIe transfers and host DRAM access. 
\namePaper{} eliminates these transfers by executing GEMM inside the SSD. 

Fifth, \namePaper{}'s energy reduction over the IFP baselines is smaller on GEMM-intensive workloads than on convolution-intensive workloads: 1.8$\times$ vs.\ 6.3$\times$ over \baseline{}, and 1.3$\times$ vs.\ 1.7$\times$ over \venice{}. 
Because all IFP systems perform the same in-flash computation, \namePaper{} reduces communication and aggregation energy. 
\namePaper{}'s lower energy reduction is because communication and aggregation account for a smaller fraction of the total energy consumption on GEMM-intensive workloads than on convolution-intensive workloads.

We conclude that \namePaper{} significantly reduces the energy consumption of distributed IFP by reducing data movement and controller involvement, and by aggregating partial results in the network.

\subsection{Sensitivity Analysis \label{subsec:sensitivity}}

\begin{figure}[b]
    \centering
    \includegraphics[width=\linewidth]{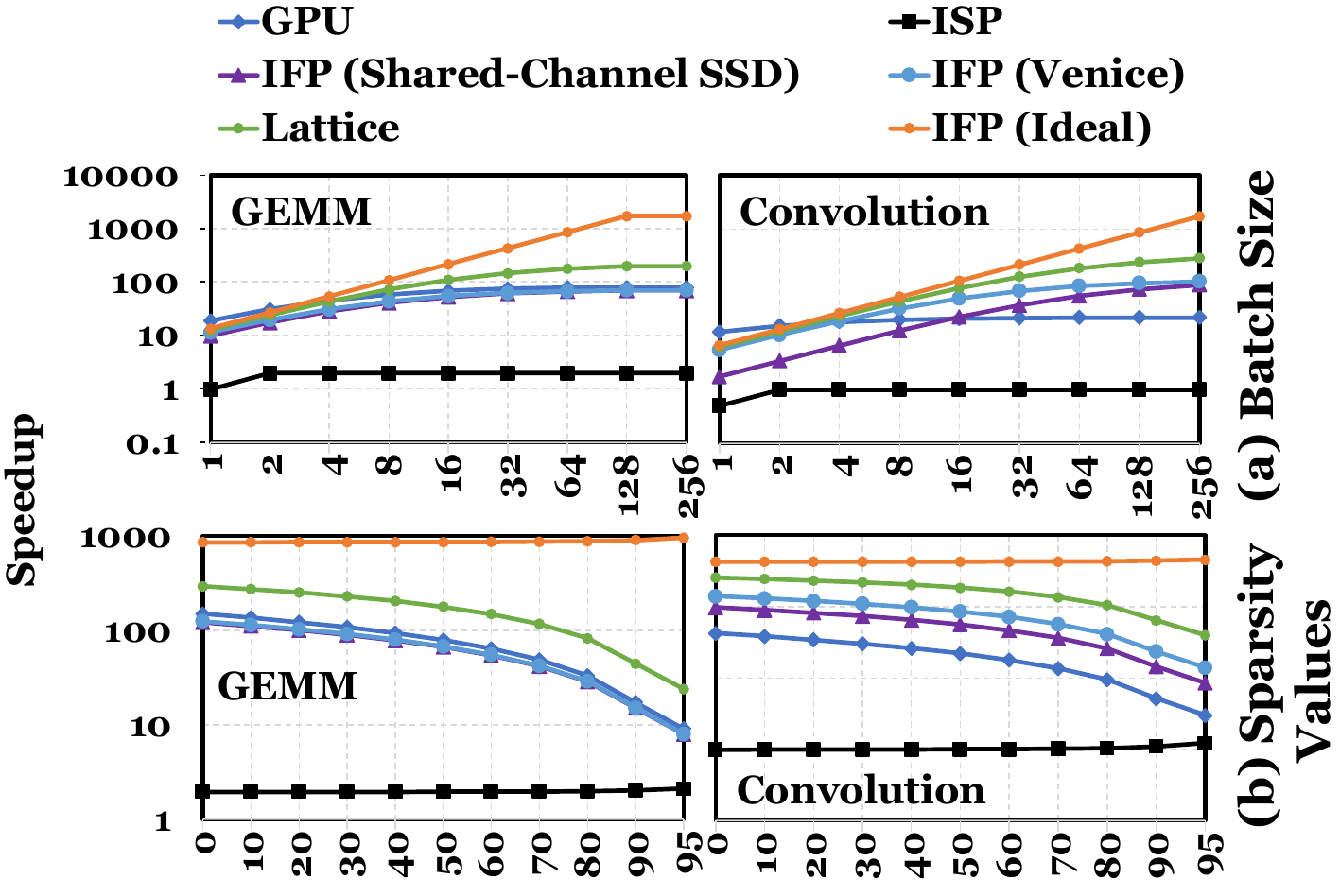}
    % \vspace{-2em}
    \caption{Sensitivity of \namePaper{} to (a) different batch sizes and (b) sparsity (\%). All values are normalized to CPU.}
    \label{fig:sensitivity_batch_sparsity}
\end{figure}

\head{Sensitivity to Batch Size}
\fig~\ref{fig:sensitivity_batch_sparsity}(a) shows the average speedup of \namePaper{} and baselines over \cpu{} on two representative workloads, GPT-OSS (GEMM) and U-Net Encoder (convolution), across different batch sizes with 50\% sparsity. 
We make four observations.
First, \namePaper{}'s speedup increases with batch size because larger batches increase activation-vector length and network-level parallelism. 
Second, the GPU's performance improves with batch size (1 to 32), because larger batches increase arithmetic intensity and tensor-core utilization.
Third, ISP's performance remains flat across batch sizes, because the weak SSD controller cores and the shared-channel bandwidth limit its computation.
Fourth, the performance of \baseline{} and \venice{} improves slightly with batch size, but saturates at a batch size of 32 due to controller-driven aggregation.

\head{Sensitivity to Sparsity}
\fig~\ref{fig:sensitivity_batch_sparsity}(b) shows the average speedup of \namePaper{} and baselines over \cpu{} on GPT-OSS (GEMM) and U-Net Encoder (convolution), for different sparsity values (0\% to 95\%) at a batch size of 64. 
We make three observations.
First, \namePaper{} benefits when sparsity increases, because routers suppress zero-valued activations during forwarding, and in-flash bit-serial arithmetic skips zero data. 
However, \namePaper{}'s speedup over CPU decreases as sparsity increases, because the normalization factor, CPU's performance, improves significantly with sparsity. 
Second, the GPU's performance improves only marginally with sparsity because the GPU loads all the data from the SSD at least once and cannot drop zero activations early.
Third, ISP's performance remains largely unchanged because the ARM cores lack support for vectorized sparse execution. 

\begin{figure}[b]
    \centering
    \includegraphics[width=\linewidth]{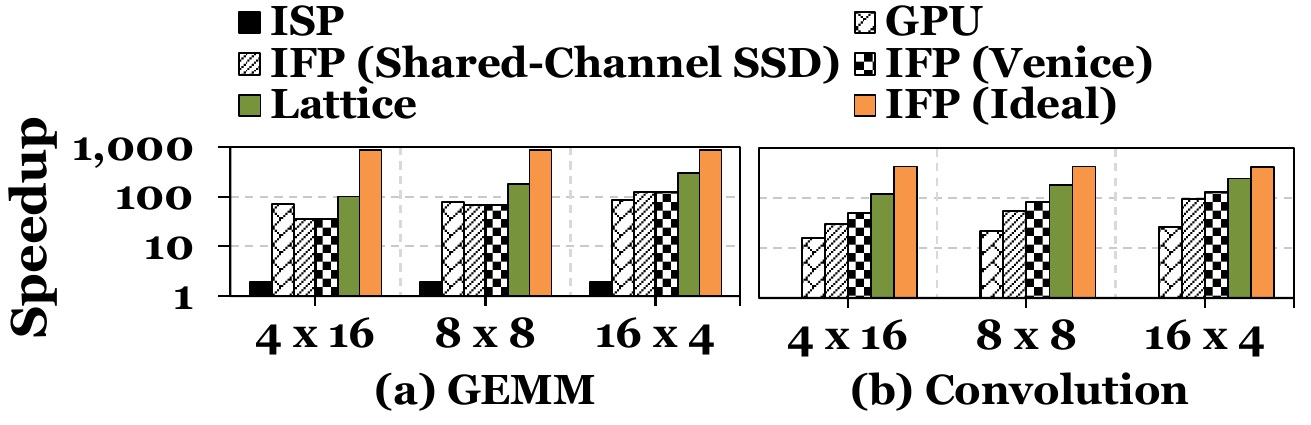}
    % \vspace{-2em}
    \caption{Sensitivity of \namePaper{} to SSD configurations with different numbers of channels and chips per channel. All values are normalized to CPU.}
    \label{fig:sensitivity_ssd_config}
\end{figure}

\head{Sensitivity to SSD Configuration}
We compare \namePaper{} and baselines on three SSD configurations with the same total number of flash chips: (1) 4 $\times$ 16 (4 flash controllers (FCs), 16 flash chips per FC), (2) 8 $\times$ 8, and (3) 16 $\times$ 4.
\fig~\ref{fig:sensitivity_ssd_config} shows the average speedup over \cpu{} on GPT-OSS (GEMM) and U-Net Encoder (convolution). 
We make two key observations.
First, \namePaper{} is less sensitive to SSD configuration than the shared-channel baselines, and its improvement is consistently higher, because its direct chip-to-chip transfer avoids the shared-channel bottleneck.
Second, shared-channel baselines (i.e., ISP, \baseline{}, GPU, and CPU) exhibit significant performance variation with channel count because internal transfer parallelism increases with channel count. 

We conclude that \namePaper{} provides its benefits across different batch sizes, sparsity levels, and SSD configurations.

\section{Discussion \label{sec:discussion}}
\head{Applicability to Other Workloads} We use ML workloads as a case study, but \namePaper{} can support a broad set of workloads that exhibit three properties: (1) distributed vector–scalar computation, (2) repeated partial-result aggregation, and (3) structured communication across data partitions.
Graph kernels~\cite{shim2022gp3d, zhang2018graphp, ahn-isca-2015, zhuo-micro-2019, huang2020heterogeneous, song2018graphr, dai-tcad-2019, giannoula-pomacs-2022} (e.g., BFS, PageRank, SpMV) exchange partial updates and perform reductions across partitions. 
\namePaper{} can perform these reductions with in-network aggregation.
Database operations (e.g., scans, filters, group-by) can benefit from \namePaper{}'s broadcast and reduction primitives, and from the routers' masking units for predicate filtering (see \S\ref{subsec:lattice_router}). 
Scientific computing workloads (e.g., stencil~\cite{singh2020nero, denzler2021casper} and finite-difference) require structured neighborhood communication across multidimensional data grids.
This communication naturally maps to the horizontal and vertical links of \namePaper{}'s 2-D mesh.
These properties appear in many storage-bound workloads, which makes \namePaper{} applicable beyond ML pipelines.
We leave the evaluation of \namePaper{} on these workloads to future work.

\head{Applicability to Other Network Topologies} 
In addition to a 2-D mesh, \namePaper{} supports other network topologies (e.g., ring, tree) based on the application's communication patterns. 
For example, tree topologies reduce latency for reduction-dominated workloads.
A ring topology (e.g.,~\cite{schuetz2007hyperlink}) reduces wiring complexity and silicon area while enabling chip-to-chip forwarding, multicast propagation, and reduction operations.
\namePaper{}'s communication substrate can support these topologies with minimal changes to the router datapath.
\section{Related Work}
To our knowledge, \namePaper{} is the first IFP-aware SSD architecture with direct chip-to-chip communication and in-network aggregation of partial results.
\S\ref{sec:results} compares \namePaper{} quantitatively with the host CPU and GPU, ISP, \baseline{}, and \venice{}~\cite{nadig2023venice}.
In this section, we briefly review prior storage-centric works and networked SSD architectures.

\head{In-Storage Processing} Prior works (e.g., \cite{ahn2022embeddedaccelerator, choe2016ndpdiffml, kim-fast-2021, niu2024flashgnn, okafor2023fusing, jang2024smart, kaplan2018prins}) place computation units near the SSD controller to accelerate neural networks. 
SmartSSD~\cite{kang-msst-2013,lee-ieeecal-2020, park2014query} integrates an FPGA in the SSD. 
These designs reduce PCIe transfers, but they do not reduce data movement across flash channels.
In contrast, \namePaper{} computes on operands in the flash chips, and it aggregates partial results without orchestration by the SSD controller.

\head{In-Flash Processing} 
Analog IFP works (e.g., \cite{kang-tc-2021, lue-iedm-2019, choi-iscas-2020, shim2020architectural, lee2022flashpim, han-aisy-2021, wang2019three, bavandpour2020mixed, lee2020neuromorphic, shim2021technological, bavandpour2021acortex, zhao2021cim, jang2025dissecting}) perform MAC operations using current summation in flash arrays, but incur high power due to analog-to-digital converters and precise current sensing.
Digital IFP techniques (e.g., \cite{park2022flash,chen2024aresflash,kim2025crossbit, chen2024search, wong2024tcam, gao-micro-2021, chen2025reis, jung2026kif}) use the operating principles of flash cells and the flash chip's latching circuitry to perform bitwise operations and bit-serial arithmetic.
However, they treat each flash chip as an isolated unit, so the SSD controller must aggregate their partial results.
\namePaper{} overcomes this limitation with a low-power inter-chip communication substrate.

\head{Networked SSD Architectures} Prior network-in-SSD architectures (e.g.,~\cite{nadig2023venice, kim2022networked, tavakkol2012network, cui2024smartnetssd, schuetz2007hyperlink}) embed an interconnection network of flash chips within the SSD, but their goals differ from \namePaper{}'s.
Venice~\cite{nadig2023venice} and Network-on-SSD~\cite{tavakkol2012network} use a 2-D mesh topology to introduce path diversity and reduce channel contention.
Like \namePaper{}, Venice~\cite{nadig2023venice} places a router chip next to each flash chip, but the SSD controller reserves a conflict-free path, and initiates every data transfer (\S\ref{subsec:motivation_networked}).
SmartNetSSD~\cite{cui2024smartnetssd} uses a multi-path flash network to accelerate read-retry operations.
Networked SSD~\cite{kim2022networked} packetizes the flash interconnect and enables direct chip-to-chip transfers via its Omnibus topology, but uses them solely for storage I/O and garbage collection.
Decoupled SSD~\cite{kim2021decoupled} adds a router to each flash controller to form a network of flash controllers, which reduces data movement through the SSD controller, but the flash chips still connect to their flash controllers over shared channels.
However, these architectures rely on the SSD controller to orchestrate every computation and data transfer, and they lack computation-aware communication primitives and in-network aggregation.
\section{Conclusion}
We propose \namePaper{}, a decentralized IFP-aware SSD architecture that transforms an SSD from a collection of isolated NAND flash chips into a cooperative IFP substrate.
\namePaper{} alleviates two architectural bottlenecks that limit distributed IFP in an SSD: the shared-channel organization and controller-orchestrated execution.
\namePaper{} builds on the key insight that IFP scalability requires decentralizing communication, not only computation.
\namePaper{} is based on three key ideas.
First, \namePaper{} embeds a lightweight IFP-aware router adjacent to each flash chip to enable direct chip-to-chip communication without the SSD controller's orchestration.
Second, \namePaper{} introduces lightweight IFP-aware communication primitives (e.g., broadcast) that support operand distribution, synchronization, and in-network aggregation of partial results.
Third, \namePaper{} performs complex arithmetic (e.g., matrix multiplication) within the flash chips and simple computations (e.g., comparison) in the routers during data transfer.
On average across five GEMM- (convolution-) intensive workloads, \namePaper{} (1) outperforms IFP on the state-of-the-art network-in-SSD architecture, Venice~\cite{nadig2023venice}, by 2.6$\times$ (2.5$\times$), and (2) reduces energy consumption by 1.3$\times$ (1.7$\times$).
\namePaper{}'s benefits come with low area and power overheads.

%%%%%%% -- PAPER CONTENT ENDS -- %%%%%%%%

%%%%%%%%% -- BIB STYLE AND FILE -- %%%%%%%%
\bibliographystyle{unsrt}
% \bibliography{refs, safari}

\end{document}